\documentclass[sigconf,authorversion]{acmart}

\AtBeginDocument{%
  }

\usepackage{enumitem}
\setlist[enumerate]{label*=\arabic*.}

\usepackage{soul}       
\usepackage{color}      
\usepackage{xspace}     
\usepackage{multirow}
\usepackage{listings}
\usepackage{subcaption}
\usepackage{framed}

\newcommand{\sys}{\textsc{Plume}}
  
\newenvironment{considerationbox}{\definecolor{shadecolor}{HTML}{F7F8F9}\begin{shaded}\noindent}{\end{shaded}}

\newcommand{\rrr}[1]{{#1}}

\copyrightyear{2025}
\acmYear{2025}
\setcopyright{cc}
\setcctype{by}
\acmConference[CHI '25]{CHI Conference on Human Factors in Computing Systems}{April 26-May 1, 2025}{Yokohama, Japan}
\acmBooktitle{CHI Conference on Human Factors in Computing Systems (CHI '25), April 26-May 1, 2025, Yokohama, Japan}\acmDOI{10.1145/3706598.3713580}
\acmISBN{979-8-4007-1394-1/25/04}

\begin{document}
\title{\sys: Scaffolding Text Composition in Dashboards}

\author{Maxim Lisnic}
\email{maxim.lisnic@utah.edu}
\affiliation{%
  \institution{University of Utah}
  \city{Salt Lake City}
  \state{Utah}
  \country{USA}
}

\author{Vidya Setlur}
\email{vsetlur@tableau.com}
\affiliation{%
  \institution{Tableau Research}
  \city{Palo Alto}
  \state{California}
  \country{USA}
}

\author{Nicole Sultanum}
\email{nsultanum@tableau.com}
\affiliation{%
  \institution{Tableau Research}
  \city{Seattle}
  \state{Washington}
  \country{USA}
}

\renewcommand{\shortauthors}{Lisnic et al.}

\begin{abstract}
Text in dashboards plays multiple critical roles, including providing context, offering insights, guiding interactions, and summarizing key information. Despite its importance, most dashboarding tools focus on visualizations and offer limited support for text authoring. To address this gap, we developed \sys, a system to help authors craft effective dashboard text. Through a formative review of exemplar dashboards, we created a typology of text parameters and articulated the relationship between visual placement and semantic connections, which informed \sys's design. \sys~employs large language models (LLMs) to generate contextually appropriate content and provides guidelines for writing clear, readable text. A preliminary evaluation with 12 dashboard authors explored how assisted text authoring integrates into workflows, revealing strengths and limitations of LLM-generated text and the value of our \textit{human-in-the-loop} approach. Our findings suggest opportunities to improve dashboard authoring tools by better supporting the diverse roles that text plays in conveying insights.

\end{abstract}



\begin{teaserfigure}
  \includegraphics[width=\textwidth]{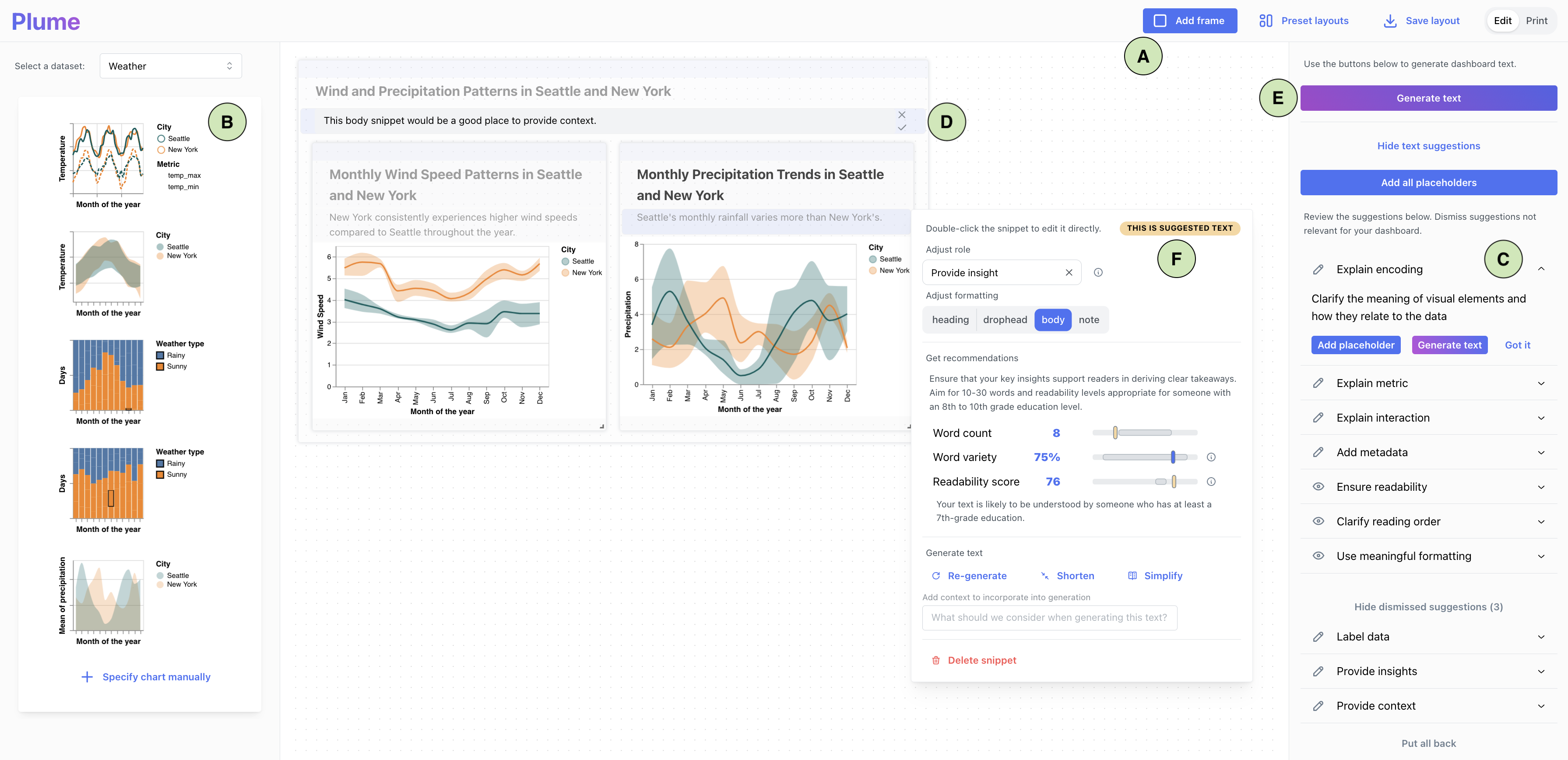}
  \caption{\sys~user interface. The user starts by adding frames to the canvas (A), dragging in charts from the charts drawer (B), and then editing text by following text suggestions in the sidebar (C). The dashboard can be further modified by dragging and resizing frames in the canvas. The figure shows the main modes of authoring text: accepting a system-suggested placeholder and typing text manually (D), asking the system to generate text based on the dashboard automatically using large language models (LLMs) by pressing the generation button (E). The user can also click into a text fragment to reveal a dropdown menu (F) with guidance on how to compose effective text, linguistic metrics of readability, and options to regenerate or refine the text using LLMs.}
  \Description{Screenshot of a user interface. In the center is a dashboard canvas with visualizations and texts. To the sides are vertical sidebars with controls. Overlaid are circles with letters A through F.}
  \label{fig:teaser}
\end{teaserfigure}

\maketitle

\section{Introduction}
Dashboards have become a ubiquitous tool in data analysis and reporting, widely used across various industries to present and interpret data visually. They help users quickly grasp complex information through charts, graphs, and other visual elements. Although visual components are vital to dashboards, the accompanying text plays a crucial role in providing context, explanations, and insights that visuals alone cannot convey. Text in dashboards serves multiple semantic roles: it provides labels and annotations, summarizes key findings, offers detailed insights, and guides users through interactions with the visualizations.

Despite text being an indispensable component of visualization dashboards~\cite{sultanum_instruction_2024}, most popular visualization and dashboard tools lack features to aid authors in composing text. This deficiency is particularly notable when compared to the robust support available for creating visualizations. For instance, systems that aid in chart design offer reasonable defaults, common chart options, and guidance on best practices for visual communication. This support spans a wide range of tools, from programming libraries like ggplot2~\cite{wickham_ggplot_2016} to the drag-and-drop interface of Tableau.

\rrr{Previous research has highlighted that the process of composing and updating text in dashboards is seen as tedious and unstructured. Dashboard authors frequently need to manually rewrite text after data updates{~\cite{zhang_understanding_2022}} or adjust it to fit different presentation contexts{~\cite{sarikaya_what_2019}}. Prior work has also found that authors want a more structured and consistent approach to writing dashboard text{~\cite{elias_annotating_2012}}, a need that arises from the wide variety of text types and goals that appear in dashboards{~\cite{sultanum_instruction_2024}}. Despite these challenges, current tools offer little to no support for text composition.} To address this gap, we developed \sys, a system designed to help dashboard authors write text for their visualizations by considering the diverse roles that text plays. \sys~supports clear organization and reading order, provides guidance for meaningful placement, and facilitates the composition of easy-to-read text, while also offering functionality to automatically generate or update text for the entire dashboard.

Our contributions are threefold. First, we conducted a formative analysis of individual text fragments in visualization dashboards to identify key features important for scaffolding and generating dashboard text. Building on prior work that analyzed types of text in dashboards~\cite{sultanum_instruction_2024}, we performed a more granular analysis to identify the common semantic roles of text---such as labels, summaries, insights, and interaction guides---and their formatting and syntactical features. Second, we conceptualized the dashboard layout as a \emph{scope dependency tree}, arguing that dashboard's layout and text are highly semantically dependent. Therefore, understanding dashboard regions in terms of their dependency relationships is crucial for both maintaining a clear reading order and for creating a system that suggests and generates relevant text automatically. Third, we designed and evaluated \sys, a system that integrates findings from prior research and our formative analysis, proposing an assisted text composition workflow for dashboard authors using generative AI. Feedback from 12 participants revealed key insights into \sys's usability and the utility of its \textit{human-in-the-loop} approach. Participants valued the flexibility and control the tool offers, such as adjusting text hierarchy and readability, while also suggesting greater customization in text placement and dynamic updates based on data changes. They appreciated \sys~for providing valuable starting points, which helped save time, while also sharing considerations about the verbosity and trustworthiness of AI-generated text, which are particularly relevant in contexts for critical decision-making. 

Our work highlights the importance of future research in supporting text authoring at different semantic levels, while ensuring coherence and consistency throughout various artifacts of data communication beyond dashboards. In addition, the research identifies implications for trade-offs between manual and AI-generated text, aiming to strike a balance between automation and user control for effective and accurate data communication.
\section{Related Work}

Our work is informed by contributions in the field regarding the role of text in dashboards and charts, text readability, and writing guidance.

\subsection{Text in Charts and Dashboards}
Integrating text in dashboards is crucial for enhancing user comprehension and interaction, serving as both instructional and interpretative elements~\cite{ottley2019curious, Hearst2023ShowIO}. Sultanum and Setlur~\cite{sultanum_instruction_2024} explored the functional and semantic roles of text in interactive dashboards, emphasizing that text not only guides users in understanding visualizations but also facilitates insight extraction. Their study highlights the diverse roles text can play, from simple labels to complex narrative elements that provide context and interpretation. Our formative work builds this categorization, offering a more granular typology of text fragments and explicitly distinguishing between semantic, typographical, and layout features of dashboard text.

The design and implementation of dashboards have been widely studied, with Bach et al.~\cite{bach_dashboard_2022} identifying common design patterns that incorporate text effectively. They provide a comprehensive analysis of how text is used in various dashboard types to achieve specific design goals, such as enhancing readability and guiding user interaction. Similarly, Sarikaya et al.~\cite{sarikaya_what_2019} investigated the broad scope of dashboard use, characterizing it by design goals and interaction levels. They proposed a framework that emphasizes the importance of text in supporting dashboard functionality and user engagement.

In the context of cooperative dashboard design, Setlur et al.~\cite{setlur_heuristics_2023} proposed heuristics that underscore the role of text, among other dashboard elements, in facilitating collaborative decision-making.  Their research suggests that well-integrated text can enhance the usability of dashboards by providing clear and concise information that supports group discussions and consensus-building. These heuristics were later repurposed by Sultanum and Setlur~\cite{sultanum_instruction_2024} to focus specifically on best practices for text use. \rrr{Srinivasan et al.}~\cite{srinivasan_dashboard_2024} complemented these perspectives by conducting a scalable analysis of dashboard designs in the wild using a large dataset from Tableau Public. Their findings reveal common patterns in how text is structured and presented, providing insights into effective text integration strategies.

The balance between text and charts is critical to dashboard design. Stokes et al.~\cite{stokes_striking_2022} examined reader preferences when integrating text and charts, finding that users prefer a harmonious balance that allows for both immediate data comprehension and deeper narrative exploration. Their study underscores the importance of strategically placing text to complement visual elements without overwhelming users.

Collectively, the heuristics and evaluations to better understand the role of text in visual analysis helped inform the development of \sys, which aims to scaffold text composition in dashboards by leveraging insights from the effective integration of text.

\subsection{Text Readability and Writing Guidance}

The importance of syntactic and lexical complexity in understanding text readability has been extensively studied in the context of educational and media texts. Edmonds~\cite{Edmonds1999SyntacticMO} explored syntactic measures of complexity, proposing metrics that help evaluate the cognitive load required to comprehend texts. This foundational work underscores the significance of syntax in assessing readability, a theme further developed by Frantz et al.~\cite{frantz:2015}, who argued for the inclusion of syntactic complexity as a distinct component in models of text complexity, highlighting its impact on reading comprehension. More recently, these findings have also been echoed by visualization researchers, with Stokes and Hearst arguing that text readability should be a major concern when integrating text and visuals~\cite{stokes2024give}.

In recent years, machine learning techniques have been employed to enhance the assessment of text readability. Laputenko et al.~\cite{Laputenko:2023} demonstrated the application of machine learning methods, such as neural networks and regression techniques, to evaluate the readability of media texts. Their approach emphasizes the role of linguistic features in determining text complexity, aligning with the goals of intelligent tutoring systems. Similarly, Greenfield and McNamara~\cite{greenfield:2008} introduced cognitively based indices for assessing text readability, which consider both syntactic and semantic features to provide a comprehensive evaluation of text difficulty.

Kurdi~\cite{kurdi:2020} focused on the classification of text complexity using linguistic information, particularly in the context of English as a Second Language (ESL) learners. His research highlighted the need for intelligent systems to offer tailored guidance based on the linguistic capabilities of learners, suggesting that such systems could significantly enhance the tutoring process. Ayadi~\cite{Ayadi:2023} investigated lexical sophistication and syntactic complexity as predictors of academic writing performance. The study concluded that lexical sophistication is a stronger indicator of writing quality than grammatical complexity, providing valuable insights for improving academic writing. Vahrusheva et al.~\cite{vahrusheva:2023} revisited the assessment of text complexity by examining fluctuations in lexical and syntactic parameters, further emphasizing the dynamic nature of text comprehension and its implications for readability assessment.

Wang et al.~\cite{wang:2024} explored linguistic features in narrative and opinion genres, analyzing their relationships with writing quality in fourth-grade writing. Their findings highlighted the critical role of linguistic features in shaping writing quality, suggesting that these features should be considered in educational tools aimed at enhancing writing skills. In the context of text readability and writing guidance, Lin et al.~\cite{lin:2008} proposed a method for measuring text readability through lexical relations retrieved from WordNet. This approach underscores the importance of semantic networks in assessing text difficulty and offers a framework for developing tools that support effective writing.

This research collectively informs the design and development of \sys, a system that scaffolds text composition in dashboards by leveraging syntactic and lexical insights to enhance readability and writing quality. 

\subsection{Text and Dashboard Generation}

Text generation for data visualizations builds on a rich body of work by integrating natural language into visual analysis workflows. Previous research explored the affordances of natural language during visual analysis to make visual analysis more accessible~\cite{srinivasan2017natural,sun2010articulate,gao2015datatone,setlur2016eviza}. Language pragmatics~\cite{hoque2017applying} and question-answering models~\cite{dhamdhere2017analyza,kim2020answering} have further demonstrated the potential of conversational models in guiding users through visual analysis tasks.

Previous research systems have shown how generated text can be used to describe interesting trends and patterns in data~\cite{mittal1998describing,srinivasan2018augmenting,kanthara2022chart,chen2020figure}. Although tools such as Tableau's Summary Card~\cite{tableausummary} and Power BI~\cite{powerbi} include basic natural language captions that summarize visual insights, \sys{} further extends these text descriptions by enabling more granular text generation for various roles, such as annotations, contextual description, insights, and interaction explanation. A related body of work has explored generating text for visualizations for accessibility purposes, including chart descriptions~\cite{Alam2023SeeChartEA, Srinivasan2023AzimuthDA} and insights-driven summaries~\cite{Obeid2020CharttoTextGN}. 

\sys{} is also aligned with the growing field of data storytelling, which highlights the importance of weaving together data, text, and visuals for compelling communication. Segel and Heer explain the value of data storytelling to make insights more memorable and persuasive~\cite{segel2010narrative}. Research indicates that narrative-driven dashboards integrate both text and charts to enhance information synthesis, addressing a known challenge in data storytelling, particularly when text and visual elements are spatially separated~\cite{kosara2013storytelling,lee2015more}. This integration allows for a more cohesive and engaging presentation of data, where textual explanations complement visualizations, helping readers to understand insights more effectively and retain key information. Systems such as Kori~\cite{latif2019authoring, latif2021kori} and VizFlow~\cite{sultanum2021leveraging} have previously explored explicit linkages between text and visuals, emphasizing narrative sequencing and rhetorical patterns~\cite{hullman2013deeper, hullman2011visualization}. 

Building on approaches like adaptive command discovery~\cite{myers2019adaptive, furqan2017learnability, corbett2016can, voicehints}, systems can incorporate suggestions that guide users in authoring dashboard text, particularly when they are unsure where to begin. This method aligns with query suggestion techniques used in exploratory search tasks~\cite{otsuka:2012, medlar:2021}, which offer follow-on queries or reformulations based on relevance~\cite{bhatia:2011,he:2009}. 

Despite these advances, little attention has been paid to \emph{text} in the context of dashboard authoring automation. Past work has focused primarily on visual aspects, such as selecting charts and widgets \cite{pandey2022medley,Deng2022DashBotID, Srinivasan2023BOLTAN} and automating layout ~\cite{Zeng2023SemiAutomaticLA}. 
Azimuth~\cite{Srinivasan2023AzimuthDA} generates an accessible textual description of a dashboard for screen readers; however, it plays primarily a supportive role adjacent to the dashboard as opposed to being thoroughly integrated as part of the dashboard and its layout.

\sys{} extends upon this body of research by offering the first system to scaffold the text composition aspect of dashboard authoring. \sys{} does so by supporting the generation of role-specific text tailored for different elements of the dashboard, including annotations, contextual descriptions, insights, and interaction explanations. By employing a hierarchy of role-specific text elements, the system uses LLMs with human-in-the-loop oversight to generate text targeted for specific communication goals during dashboard authoring.

\section{Usage Scenario}

Consider Skye, a data analyst tasked with creating a visualization dashboard to compare weather patterns in Seattle and New York---two of their firm’s primary client locations. After understanding the client's needs and crafting visualizations that report on several relevant weather aspects of the two cities, they begin assembling a dashboard that will help users assess differences between the two cities.

\begin{figure*}[t]
    \centering
    \includegraphics[width=\textwidth]{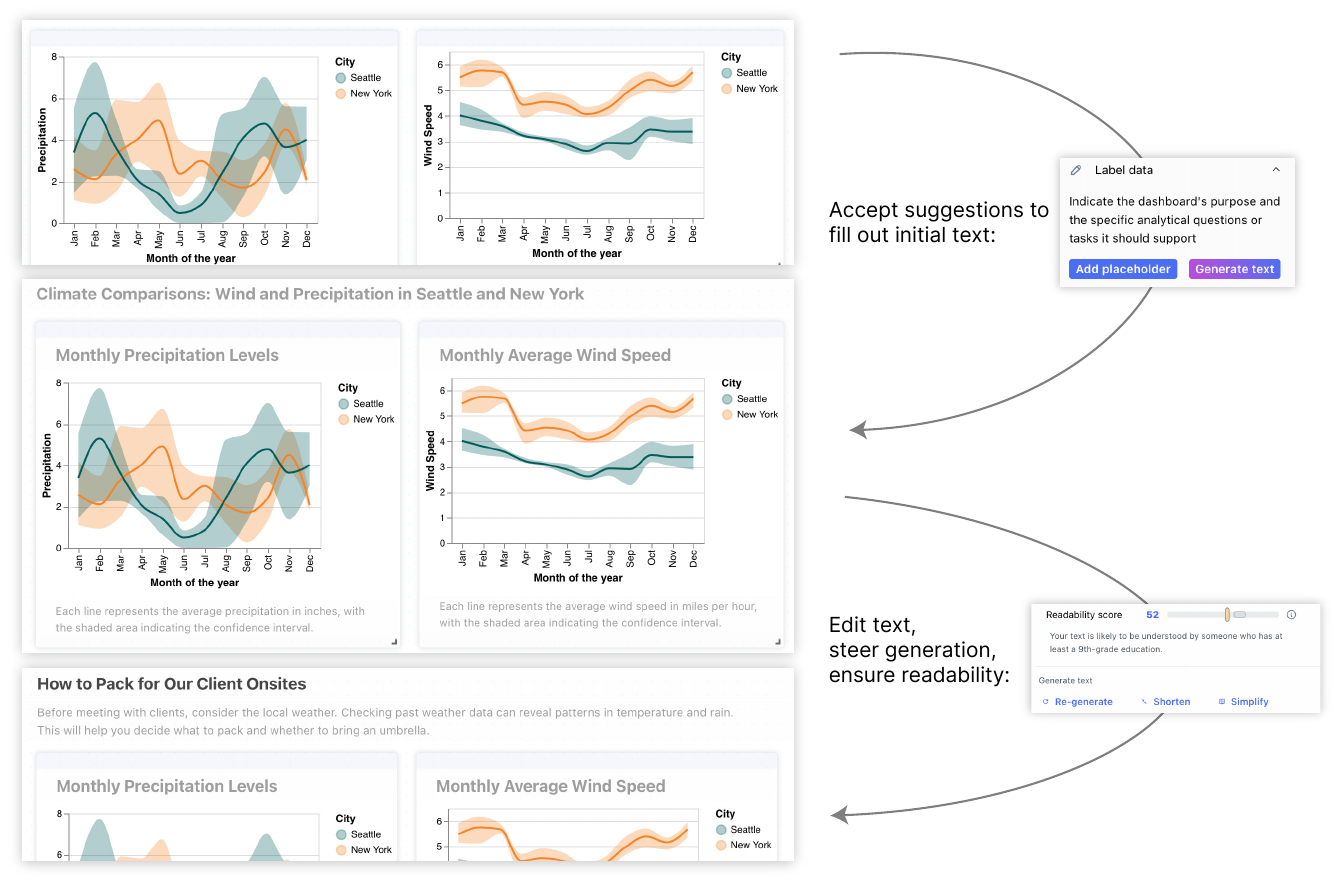}
    \caption{An illustration of the usage scenario. An analyst starts by laying out the dashboard and adding charts to the canvas. They may then accept \sys~suggestions to fill out the initial text based on the text roles they deem necessary. The analyst then makes final edits to the text, generates more accurate text using \sys~now that it is aware of the analyst's intended tone and goals, and finally uses linguistic metrics in \sys~to ensure readability.}
    \Description{Screenshots of a dashboard with two line charts, first without any text, then with generated titles and encoding description, and lastly with a context-specific title and short paragraph.}
    \label{fig:usage_scenario}
\end{figure*}

First, they configure the dashboard layout using \sys{}, defining various frames that will each contain specific visualizations or nested subsections.
With the layout established and charts placed, Skye proceeds to the text authoring phase. \sys~provides a panel with text recommendations with best practices for effective dashboard text composition. Skye reviews each suggestion sequentially. The initial recommendation---seen as the first step in Figure~\ref{fig:usage_scenario}---advises labeling the sections of the dashboard by adding titles. By clicking the \emph{Generate text} button, Skye allows the system to automatically generate and place text based on the visualizations on the canvas, positioning it at the top of each section and formatting it in large, bold letters. The subsequent suggestion instructs Skye to incorporate insights into the dashboard. Opting for a more exploratory approach, Skye chooses to bypass this suggestion for the time being.

Skye then accepts another recommendation from \sys~to clarify the visual elements of the visualizations. Upon acceptance, the system inserts explanatory notes under each visualization, detailing how to interpret the charts. This is particularly useful for a line chart featuring shaded areas---created by a previous team member---where Skye was unsure of the calculations. \sys~deduces from the visualization specifications that these shaded areas represent 95\% confidence intervals and includes this explanation in the text.

Next, Skye addresses the suggestion to add metadata to the dashboard. Instead of generating text automatically, \sys~prompts Skye to specify the author, data source, and any disclaimers or caveats. Skye recalls that the 2024 data is incomplete and partially imputed, which is crucial information for viewers. They enter this detail into a text box, and \sys~incorporates the metadata into a straightforward English sentence at the bottom of the dashboard.

Upon reviewing the dashboard, Skye decides to refine its message. Instead of merely exploring weather patterns, they aim to create a dashboard that informs viewers about packing for client visits. Skye renames the dashboard to \emph{How to Pack for Our Client Onsites} by double clicking the generated title and making the change. By clicking the checkmark, Skye locks in the updated title. They then accept a suggestion to provide context in the sidebar, which generates a paragraph under the dashboard title. This paragraph reflects the new title and infers the dashboard’s communication goal, advising viewers to prepare for trips by, for example, packing an umbrella for Seattle’s rainy season.

A subsequent sidebar suggestion prompts Skye to evaluate the readability of the text. They access each snippet’s dropdown menu to adjust its role, formatting, and prominence. Using \sys's readability metrics, Skye notices that the context paragraph has a low readability score, indicating it may be too complex for some readers. They click the \emph{Simplify} button, prompting \sys~to rewrite the text to improve clarity. The revised text is now more accessible.

By now, Skye had performed a total of 6 clicks and written 3 phrases, yet the dashboard contains complete text guiding its audience in how to use it, including a data caveat and an encoding clarification that Skye would have otherwise forgotten to include. Skye continues to refine the dashboard, manually adjusting the language and accepting a suggestion to describe interactions with the charts. Finally, they enter print preview mode to review the dashboard without system controls. Satisfied with the result, Skye shares the dashboard with their colleagues.
\section{Design Process}

To inform the design of \sys, we first conducted a formative qualitative analysis of exemplar uses of text in dashboards and then used our findings to derive themes and considerations.

\begin{table*}
    \centering
    \includegraphics[width=0.9\textwidth]{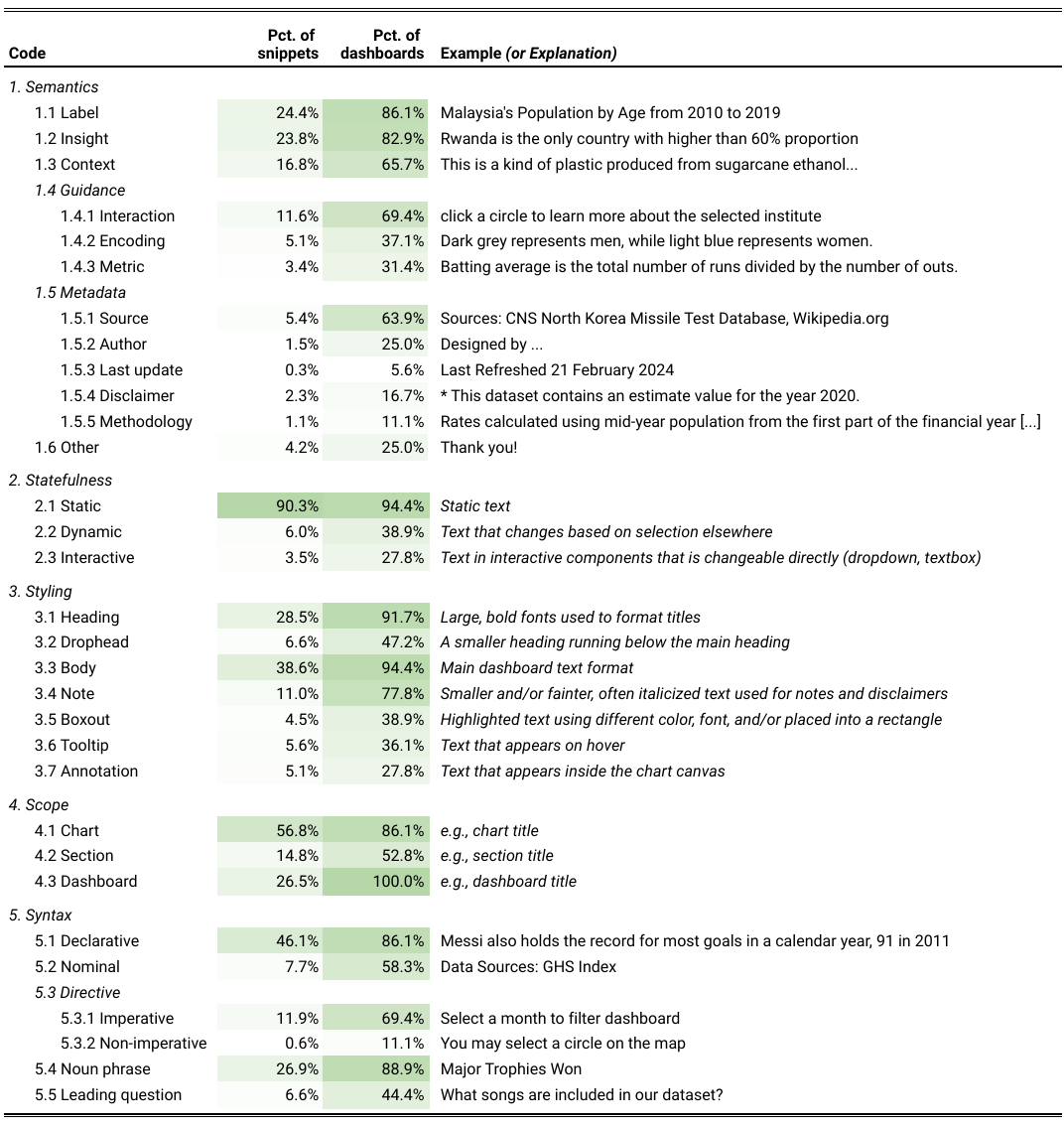}
    \caption{Final codebook version used to code all text snippets in the formative survey. Each text snippet was described by one code from each of the five sections: semantics, statefulness, styling, scope, and syntax.}
    \label{fig:codebook}
\end{table*}

\subsection{Formative Work}
\label{sec:formative}

We conducted a fine-grained survey of text in a set of exemplar dashboards to inform our tool design and identify a set of dashboard text characteristics necessary for generation.

\subsubsection{Dashboard Corpus}
We constructed a corpus of dashboards from the 190 dashboards reviewed by Sultanum and Setlur's study of text use in dashboards~\cite{sultanum_instruction_2024}. Their corpus, in turn, is composed of dashboards shared by the participants of their study, another set reviewed by prior studies on dashboard use~{\cite{bach_dashboard_2022, sarikaya_what_2019, srinivasan_dashboard_2024}}, as well as a set of highly rated public Power BI dashboards. With the goal of arriving at a set of \emph{exemplary instances} of text in dashboards rather than simply a representative sample, we chose a set of 40 dashboards that Sultanum and Setlur identified as examples of `Strong Application' of their proposed heuristics for the use of text~\cite{sultanum_instruction_2024}. We additionally ensured that the sample covers a variety of dashboard genres~\cite{bach_dashboard_2022} (13 analytic, 20 infographic, 7 other) and sources (30 Tableau Public, 8 Power BI, 2 other web visualization tools). We provide the list of reviewed dashboards in supplemental materials.

We then processed our corpus of dashboards by extracting all textual elements manually, using OCR to speed up the process. We separated the text into the smallest possible fragments of prose---phrases, sentences, or short paragraphs---that described the same idea or action and had the same formatting and layout. If a fragment was still found to encompass multiple codes in our codebook, we further split the fragment into smaller semantic chunks at a later step during the coding process to ensure a single code per fragment. We term these text fragments \textbf{snippets}, which form the basic unit of our analysis. Our corpus resulted in 672 snippets, or just under 17 snippets per dashboard, on average.

\subsubsection{Methods}

We conducted a thematic analysis of the resulting 672 snippets using the template analysis approach~\cite{king_using_2004}. Template analysis was particularly well-suited to our study for several reasons. First, it focuses on creating a hierarchical codebook of codes called the template. Since our primary goal was to identify key parameters and dimensions of dashboard text---both to suggest these parameters to users and to design descriptive prompts for future LLM generation---we aimed to construct and validate a meaningful codebook. Second, unlike grounded theory approaches, template analysis does not claim to report a ground truth set of codes that exist outside the context of our study. Instead, we had specific goals and inductively hypothesized codes that guided our analysis toward aspects of dashboard text relevant to tool design. Lastly, because we intended to contribute a labeled data set of dashboard text snippets for future research, we chose not to use a more open-ended thematic analysis and instead performed complete coding of our data set.

Following the template analysis protocol, the first author created an initial coding template based on the research goals, a review of relevant literature, and an examination of 6 of the 40 dashboards used in the analysis. The authors then met to discuss whether the codebook captured dimensions relevant to tool design and whether the codes were understandable. The first author proceeded to code 30 dashboards, iteratively updating and refining the template. Subsequently, the senior author conducted a quality check by independently coding the remaining ten dashboards, confirming that the codebook was straightforward to apply and captured all relevant dimensions. Throughout this process, all authors met daily to discuss changes. At the end, the final codebook was complete and the first author reviewed the entire sample of snippets to ensure it was coded according to the latest version of the codebook.

Table~\ref{fig:codebook} presents the final codebook used to code all the data, along with the prevalence of each code across all snippets and dashboards. Each text snippet can be described by a total of five codes, one code from each category: semantics, statefulness, styling, scope, and syntax. In contrast to previous seminal work describing types of text in dashboards by Sultanum and Setlur~\cite{sultanum_instruction_2024}, our codebook provides a more granular analysis of text. Firstly, our codebook focuses on individual fragments rather than entire blocks of text. Secondly, we offer a more fine-grained categorization of text, distinguishing, for instance, between the different subtypes of guidance. Additionally, the five categories in our codebook allow us to compare and contrast the different facets of dashboard text fragments, such as semantic goals and typographical styling, independently: one author could prefer chart titles to merely act as labels but another one could prefer titles to contain insights. Lastly, we describe dimensions of dashboard text critical to our system not mentioned in the previous work altogether, such as scope and syntax. The supplemental materials for this paper include the final coded data set, as well as an audit log and snapshots of all iterations of the codebook described above.

\begin{figure*}[h]
     \centering
 
    \begin{subfigure}[b]{.7\linewidth}
        \includegraphics[width=\textwidth]{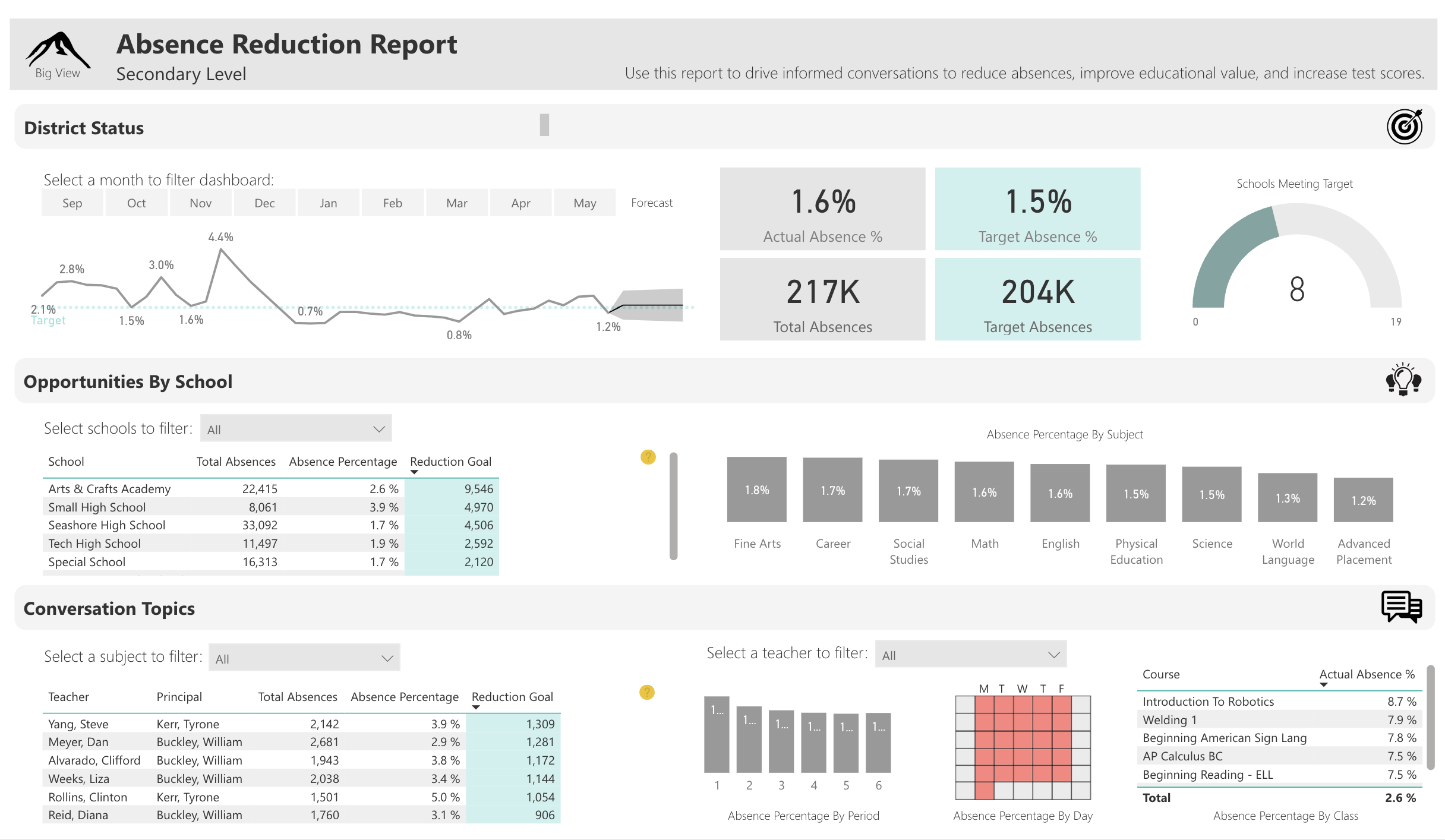}
        \caption{A dashboard tracking absences in school (\href{https://community.fabric.microsoft.com/t5/Data-Stories-Gallery/Attendance-Tracker-Improving-School-Attendance-by-Decisive-Data/td-p/136078}{source}).}
        \label{fig:example_dash_absence}
    \end{subfigure}

    \begin{subfigure}[b]{.49\linewidth}
        \includegraphics[width=\textwidth]{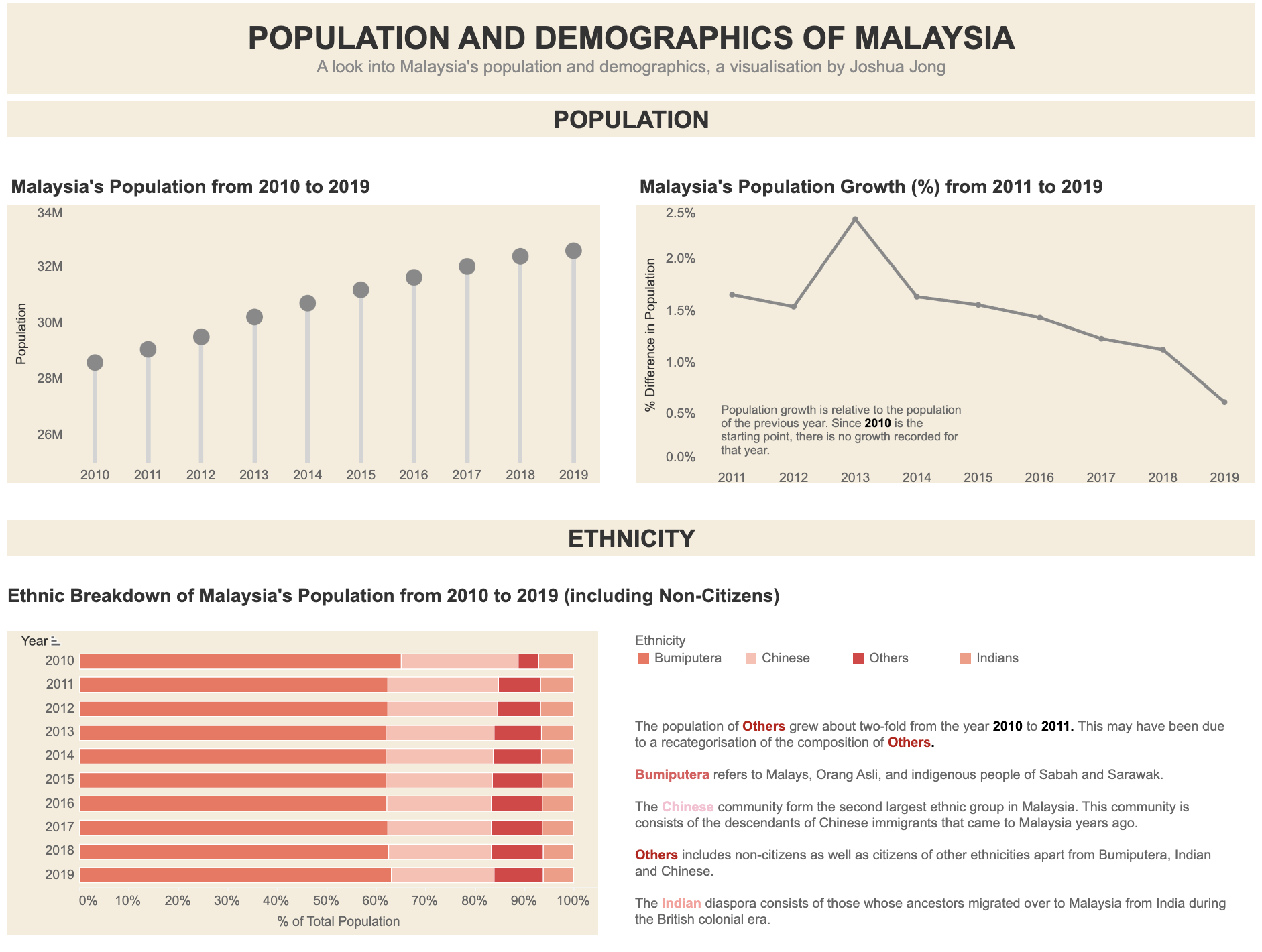}
        \caption{A dashboard visualizing demographics of Malaysia (\href{https://public.tableau.com/app/profile/joshua.jong6344/viz/assignment1_16295362252850/Dashboard13}{source}).}
        \label{fig:example_dash_malaysia}
    \end{subfigure}
    \begin{subfigure}[b]{.49\linewidth}
        \includegraphics[width=\textwidth]{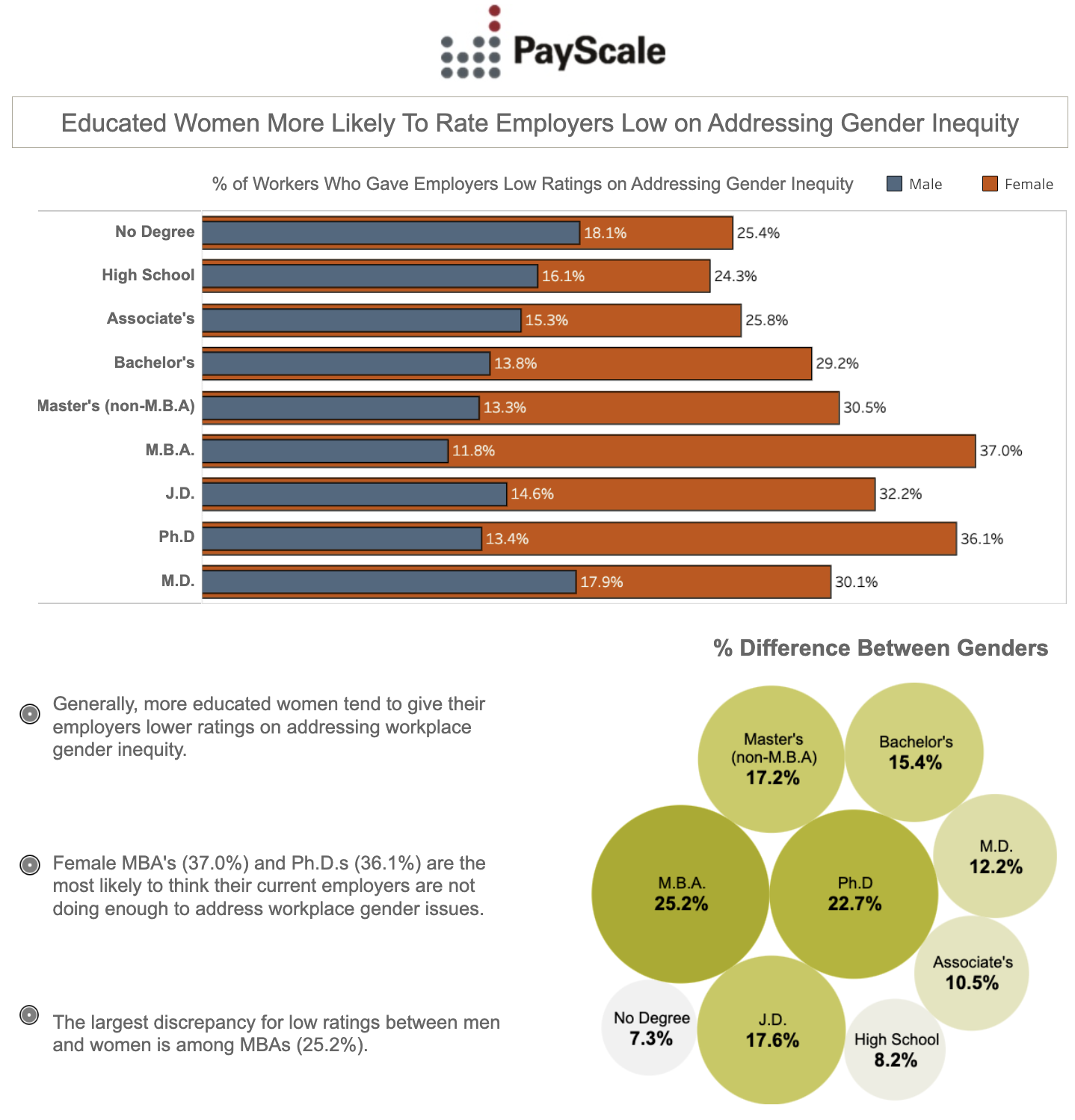}
        \caption{A dashboard analyzing the gender pay gap (\href{https://public.tableau.com/app/profile/payscale/viz/6-Education/6-Education}{source}).}
        \label{fig:example_dash_payscale}
    \end{subfigure}

    \caption{The dashboards in our formative study displayed varying levels of coordination between the visual layout of text and its semantic scope. In all cases, the Gestalt principle of proximity helps trace which part of the dashboard each text describes, though with differing clarity. Dashboard (a) uses a rigid grid, with each text placed directly above the relevant chart, section, or entire dashboard. Dashboard (b) also employs a grid but sometimes places text to the side of the chart or as an annotation on top of it, still maintaining clear associations. In contrast, Dashboard (c) places most text in its own section equidistant from two views, making it harder to visually infer the corresponding view.}
    \Description{Three screenshots of dashboards. A is an enterprise-style dashboard with tables, bar charts, and a line chart. B is a dashboard with bar charts and a line chart. C is a dashboard with a bar chart and a bubble plot.}
    \label{fig:example_dash}
    \vspace{1cm}
\end{figure*}

\subsection{Findings and Design Considerations}
\label{sec:design}

Based on our formative work, we present our findings as a set of themes that summarize our codebook and serve as design considerations for \sys. Each theme reflects an observation about a group of codes or a contrast between sets of codes from Table~\ref{fig:codebook}.

In summary, these themes highlight opportunities to provide dashboard authors with scaffolding that supports and simplifies common text practices. At the same time, each theme also reflects a tension between being too prescriptive and allowing authors complete freedom to modify their dashboard text.

\subsubsection{Types of Text}
We found that 96\% of all text snippets in our survey align with one of the identified semantic goals represented by codes 1.1--1.6, such as labeling data, explaining visuals, or surfacing disclaimers. This fact suggests that the vast majority of dashboard text serves one of these common goals. At the same time, a quarter of the dashboards included at least one snippet that did not conform to these roles, as shown by code 1.6 \textit{Other}. Such text typically included messages from the author (e.g., ``Thank you!''), served as a creative outlet, or included song lyrics and poems relevant to the data theme.

\begin{considerationbox}\textit{Design Consideration:}\\
The system should \textbf{surface common text semantic goals as options or suggestions}, provide easy ways to place and generate each of them, and offer guidance on how to author them if the user prefers to do so manually. At the same time, the system should allow the user to input any other text that may not directly fit these common goals.
\end{considerationbox}

\subsubsection{Text Placement \& Styling}
Our formative work uncovered strong correlations between the semantic goal of text (codes 1.1--1.6), its scope (codes 4.1--4.3), and its positioning and formatting (codes 3.1--3.7). For instance, as seen in dashboard examples in Figure~\ref{fig:example_dash}, data labels are typically formatted as headings and placed at the top of charts, sections, or the entire dashboard. Similarly, metadata is usually formatted in a smaller, fainter font and placed at the bottom or in the corners of the dashboard. Although such formatting and placement choices are common across text modalities, they are also influenced by the author's stylistic preferences, intended reading order, or desire to draw attention to specific text or section.

\begin{considerationbox}\textit{Design Consideration:}\\
\sys~should \textbf{select appropriate formatting and placement for text} depending on its semantic goal by default. However, since text formatting also reflects stylistic and design choices by the author, \sys~should \textbf{allow users the flexibility to modify} the placement and formatting of text.
\end{considerationbox}

\subsubsection{Text Generation}
Although all semantic goals of text coexist within the dashboard and contribute to the audience's understanding, the author draws on different information sources to craft them. For instance, when writing a data label, the author takes into account the semantics of the data, but offering context also requires incorporating world knowledge and considering the intended audience. As a result, a text generation system would similarly need to draw on different sources of information, such as raw data or visualization specification, to effectively support the author.

We identified three categories of text semantic goals in terms of generation potential and expected accuracy: text that can be accurately generated based on information readily available to the system (e.g., labels and guidance text), text that can be generated but requires approval or direction from the author (e.g., insights and context), and text that necessitates user input (e.g., disclaimers and methodology). This categorization roughly mirrors the semantic levels of text used in accessible visualizations introduced by Lundgard \& Satyanarayan~\cite{lundgard_accessible_2022}, who note that although L1 and L2 levels may be straightforward to generate, current technology often struggles at producing accurate and relevant higher-level L3 and L4 descriptions.

\begin{considerationbox}\textit{Design Consideration:}\\
LLM prompts for \textbf{different semantic goals of text would need to include the appropriate relevant information}, such as the raw data or the visualization specification. The system should avoid guessing or generating content without the necessary information and instead prompt the user to specify missing details. Additionally, \sys~should not always assume that LLM-generated text reflects the user's communication goals and should offer ways for users to steer the generation by editing, augmenting prompts, and marking certain text as generated as intended.
\end{considerationbox}

\subsubsection{Text Scope}
\label{sec:text_scope}
Codes 4.1--4.3 in the codebook indicate that each text fragment refers to either a single chart, a group of charts, or the entire dashboard. For instance, an encoding text snippet offers guidance on how to read a particular chart or set of charts, while a label can refer to a specific chart or the whole dashboard.

Text scope is primarily a semantic property; however, it is closely tied to the visual appearance. Typically, text is visually connected to the specific chart or dashboard section it references through Gestalt principles~\cite{Wertheimer1923-WERLOO}, such as proximity or enclosure. Ideally, relevant text should appear adjacent to the chart it describes and reside within the same enclosed region. This relationship is often nested, as shown in Figures~\ref{fig:example_dash_absence} and \ref{fig:example_dash_malaysia}, where the semantic scope of the text aligns clearly with the visual scope. However, the dashboard in Figure~\ref{fig:example_dash_payscale} diverges from this pattern. Insights are presented as bullet points equidistant from two views, without a clear visual cue to indicate which view the text refers to. As a result, the viewer must carefully read the text and examine both charts, making it more difficult to quickly grasp the relationship between the text and the visual data, reducing overall clarity.

\begin{considerationbox}\textit{Design Consideration:}\\
For readability and quicker comprehension, it is important that text is clearly visually related to the chart or dashboard section it describes. The system should \textbf{ensure that text scope is reflected visually} to aid in the quick understanding of references.
Additionally, the scope of text must be surfaced to the LLM generation to ensure a sensible reading order in the generated text.
\end{considerationbox}
\begin{figure*}[t]
    \centering
    \includegraphics[width=\textwidth]{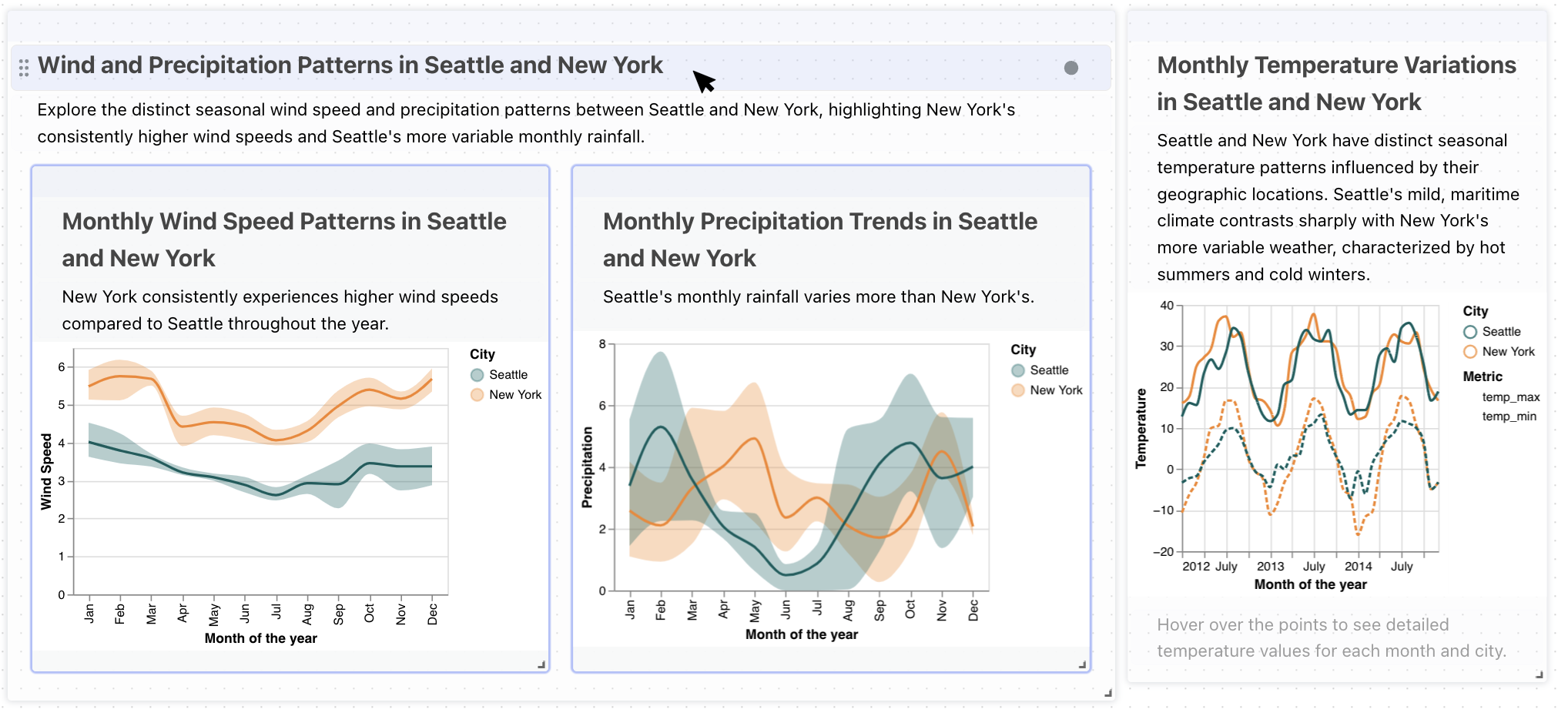}
    \caption{View from composing a dashboard in \sys. When hovering over a text fragment, \sys~highlights the descendant frames used for its text generation with a blue border. The hovered text remains independent of the sibling frame to the right.}
    \Description{A view of a dashboard with two sections. The left section is titled Wind and Precipitation Patterns in Seattle and New York. It contains two visualizations: a line chart of wind patterns and a line chart of precipitation patterns. The right section contains one chart, titled Temperature Variations in Seattle and New York.}
    \label{fig:weather_example}
\end{figure*}

\section{\sys}
At the core of the system lies the idea of \emph{dashboard frame tree}, leveraged for most features of \sys. We maintain the dependence relationships between each subsection of the dashboard as a tree structure in the system's internal state. The idea of sequentially organizing visualizations has been previously used for the purposes of identifying similar charts~\cite{kim2017graphscape} and converting notebooks to presentations~\cite{li2023notable}, however in \sys{}, this approach is primarily utilized for generating and placing semantically meaningful and coherent text. This tree structure reflects the \emph{text scope} described above in Section~\ref{sec:text_scope} and allows every text snippet to be ``aware'' of whether it needs to describe a particular chart, compare and contrast multiple charts, or summarize a dashboard section. 

In terms of workflow, {\sys} takes inspiration from the workflow offered by common dashboarding tools, such as Tableau and Power BI, where visualizations are specified ahead of time and later laid out in a dashboard and accompanied by text. This workflow both offers the advantage of familiarity to dashboard authors (as confirmed later by our evaluation in Section~{\ref{sec:study}}) and also allows us to leverage visualization specifications to generate more accurate text (as described in Section~{\ref{sec:generation}}).

In this section, we describe the system design of \sys~by tracing the importance of the dashboard tree for surfacing a coherent reading order for the author, offering suggestions for meaningful placement of text in the dashboard, as well as identifying relevant context for generating new text using LLMs.

\subsection{Dashboard Frames}

The system visibly surfaces text scope---or which part of the dashboard the text is referring to---by enforcing that every piece of text and every visualization is placed into a visually salient frame. For example, Figure~\ref{fig:weather_example} shows a screenshot of \sys~with a simple scenario of just two levels of frames. If we focus on the highlighted section of the dashboard, we can see that the inner frames contain data visualizations and text that refers exclusively to the contents of that frame---wind speed and precipitation. Meanwhile, the text in the outer frame provides a summary of its children's frames. Importantly, it does not directly depend on the text in its sibling frame to the right.

\sys~surfaces the relationship between sections of the dashboard by enforcing that all content---text or visualizations---is enclosed within rectangular frames which are visible while editing. When hovering over a piece of text with a cursor, \sys~additionally highlights direct descendants of that frame, making it particularly noticeable what the text in question is referring to. This highlighting can be seen as the blue borders in Figure~\ref{fig:weather_example}. 

By leveraging the Gestalt principles~\cite{Wertheimer1923-WERLOO}, the visually salient enclosures make it easy for the dashboard author to maintain a semantically meaningful reading order for their text and to verify that information relevant to the visualization is placed near it. Moreover, the layout also allows the author to surface the relative importance of visualizations by adjusting their depth in the dashboard tree. For instance, it would be completely reasonable for the author of the dashboard in Figure~\ref{fig:weather_example} to include the frame with the temperature visualization as a sibling frame of the other visualizations. This action would, however, also have semantic effects throughout the dashboard: the text in the parent frame would then need to describe, summarize, or contextualize all three views.

Aside from clarifying text scope for the dashboard authors and viewers, maintaining a clear dependency tree of dashboard content is crucial for \emph{placing} and \emph{generating} text by our system, as discussed below in Sections~\ref{sec:guidance} and \ref{sec:generation}, respectively.

\begin{table*}[t]
    \centering
    \includegraphics[width=0.8\textwidth]{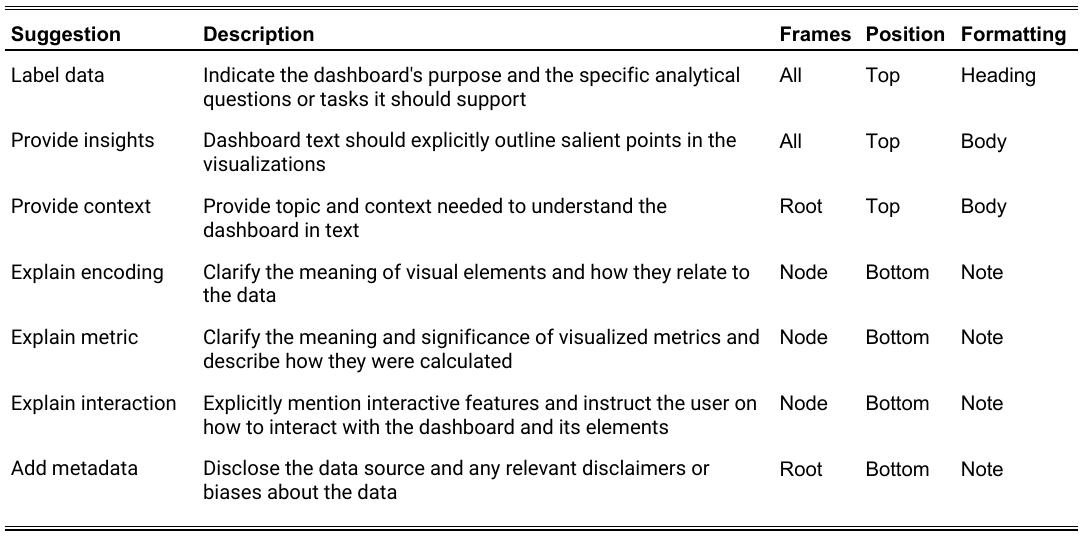}
    \caption{Global guidance suggestions and their corresponding addition of text to the dashboard upon acceptance. Each suggestion is based on one of the snippet text roles in \sys. The descriptions are also based on dashboard text heuristics from Sultanum and Setlur~\cite{sultanum_instruction_2024}.}
    \label{tab:suggestions}
\end{table*}

\subsection{Writing Guidance \& Feedback}
\label{sec:guidance}

Once the author completes the dashboard layout and adds visualizations to relevant frames, \sys~provides dynamic feedback and recommendations for authoring text. \sys~offers two distinct types of text writing guidance: global guidance and snippet-level guidance.

\subsubsection{Dashboard-level Guidance}

\sys~offers guidance for authors on how to fill out their dashboard when starting from a canvas with no text. Dashboard-level guidance consists of a series of suggestions surfaced in the sidebar (panel C in Figure~\ref{fig:teaser}) that recommend what text may be useful in the dashboard and where it should be placed. Based on our preliminary work, we have identified the seven common types of text, or \emph{text roles} that typically appear in dashboards: for instance, text that serves to label the data or text that explains how to interact with the dashboard. Each suggestion corresponds to one of the seven text roles that exist in \sys, derived from the semantic codes 1.1--1.6 in our formative codebook from Table~\ref{fig:codebook}, as well as additional reminders to consider readability, reading order, and formatting of the dashboard. These suggestions are based on dashboard text use heuristics from Sultanum and Setlur~\cite{sultanum_instruction_2024} as well as the results of our formative work outlined in Section~\ref{sec:formative}. The suggestions are not intended to be hard and fast rules but rather reasonable defaults that are as complete as possible. Consequently, any suggestions can be dismissed, and text role suggestions can additionally be accepted to generate text or text \emph{placeholders} for the dashboard. Users can also accept all suggestions at once to fill the entire dashboard.

Placeholder text provides users with a template suggestion for what type of text they should add to the dashboard and where it should be placed. Placeholders contain text that explicitly references one of the text types and instructs the user to type it in (e.g., ``This would be a good place to label your data''). An example of a placeholder instructing the user to write a context snippet can be seen in the outer frame in Figure~\ref{fig:teaser}. 

When accepted, placeholders are put in a relevant position in the frames. Dashboard frames, introduced in the previous section, play an important role in meaningfully placing new text snippets. Firstly, certain text roles may need to appear multiple times throughout the dashboard but differ in their scope. In such cases, placing text on different frame levels helps the system distinguish between a suggestion to describe an insight about an individual chart versus providing a takeaway insight for the whole dashboard. Secondly, certain types of text are only relevant for frames on one specific level of the dashboard frame tree. For instance, authorship credits typically appear only once in the outermost (or root) frame. Meanwhile, text explaining how to interact with a given chart is typically found immediately adjacent to the relevant chart (in a leaf frame). In either case, maintaining a clear relationship between the frames allows the system to place the suggestion in the most appropriate position. Table~\ref{tab:suggestions} gives an overview of the text role-based suggestions and their corresponding placeholder placements upon acceptance. 

\begin{figure}[th]
    \centering
    \includegraphics[width=0.9\linewidth]{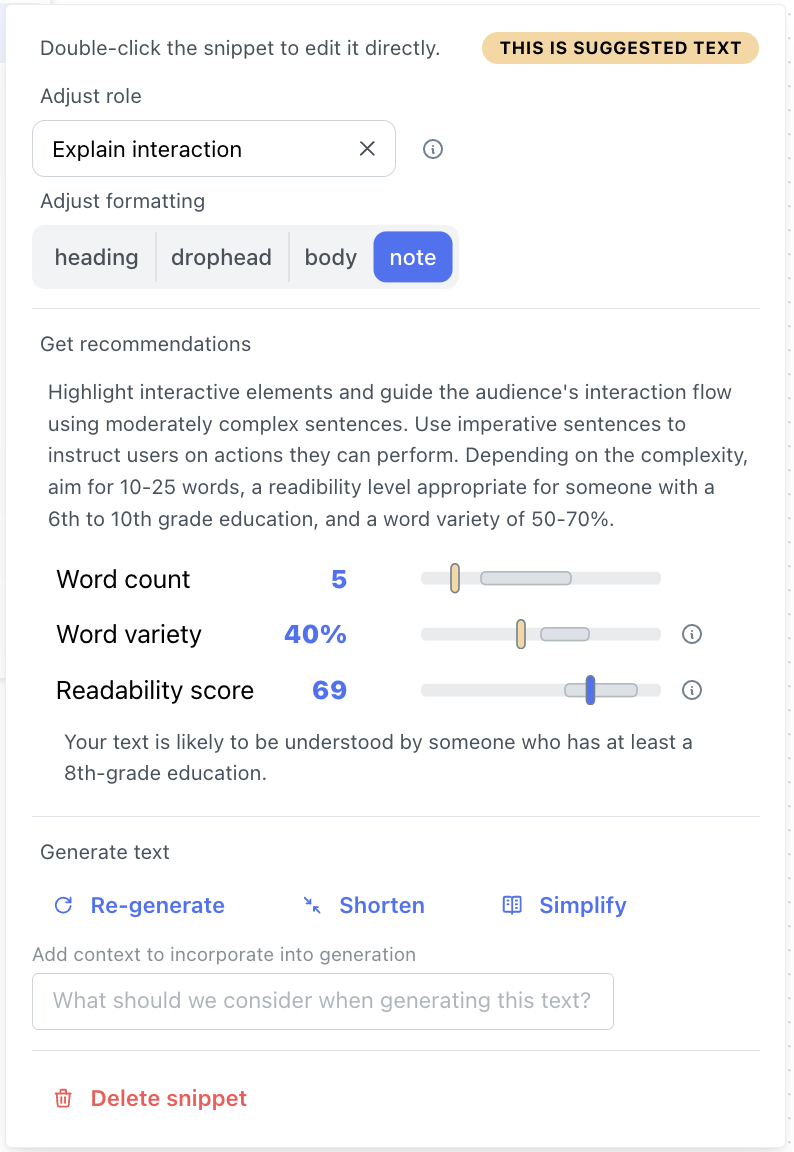}
    \caption{Snippet dropdown menu that contains snippet-level writing guidance. Based on the selected text role, \sys~provides recommendations for how to approach the writing of such text as well as surfaces text metrics. The user may then either manually rewrite the text or use \sys~to re-generate, shorten, or simplify their text to ensure that it is readable at the level appropriate for their audience.}
    \Description{An interface screenshot with buttons and toggles. The top section has dropdowns for role and formatting. The middle section has a text description of readability metrics and point plot charts of word count, word variety, and readability score. The bottom section has controls to regenerate, simplify, shorten, or delete the snippet.}
    \label{fig:dropdown}
\end{figure}

\subsubsection{Snippet-level Guidance}

We developed recommendations for writing text tailored to each of the seven text roles in the system by analyzing the numeric ranges of common text metrics, such as word count, lexical density, and readability, in the sample snippets from our formative work and drawing on related research in readability and writing guidance \cite{bird2009natural, kincaid1975derivation}. Each recommendation provides a brief overview, followed by clear guidelines on average expected word count, word variety, and readability for the text role in question, along with a visualization illustrating how the current dashboard text measures up against these benchmarks. For example, to convey analytical questions or tasks, we recommend adding concise annotation snippets that explicitly outline salient points in visualizations. As a guideline, these annotations should aim for a Flesch-Kincaid Grade Level of 8--10, with a word count of 10--20 words. We provide the complete guidance templates for each text role in the supplemental materials.

Figure~\ref{fig:dropdown} shows the dropdown menu view when a text snippet is clicked on. To measure word variety, we calculated the snippet's lexical density, measured as a percentage of content words among all words in the snippet. We define content words as those outside of NLTK's list of English stopwords~\cite{bird2009natural}. The readability score is calculated using the Flesch-Kincaid readability tests formula~\cite{kincaid1975derivation}.

\begin{figure}[tp]
    \begin{subfigure}[b]{\columnwidth}
        \centering
        \includegraphics[width=\columnwidth]{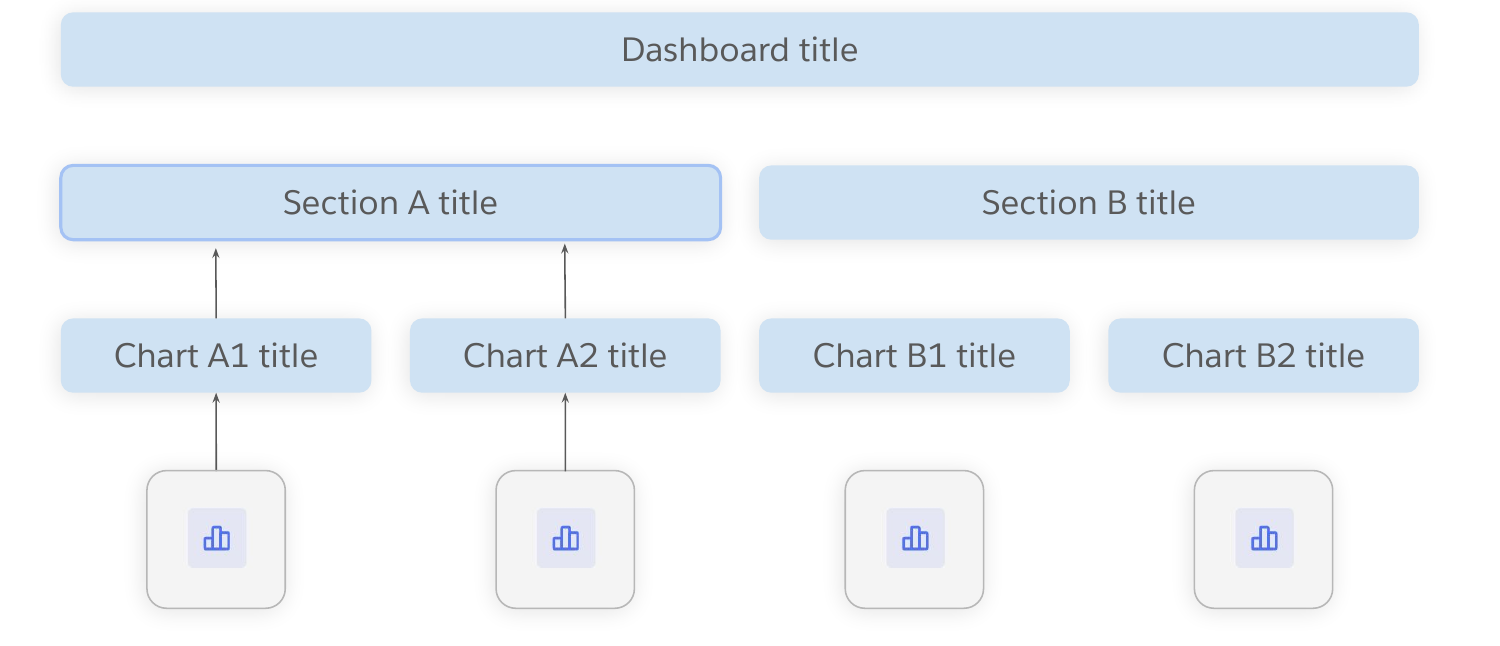}
        \caption{In a basic scenario, Section A title would be generated as the summary of its children titles, which in turn are generated from the visualization specifications. Section A title generation does not depend on any other nodes of the dashboard tree.}
        \label{fig:tree_a}
    \end{subfigure}

    \begin{subfigure}[b]{\columnwidth}
        \centering
        \includegraphics[width=\columnwidth]{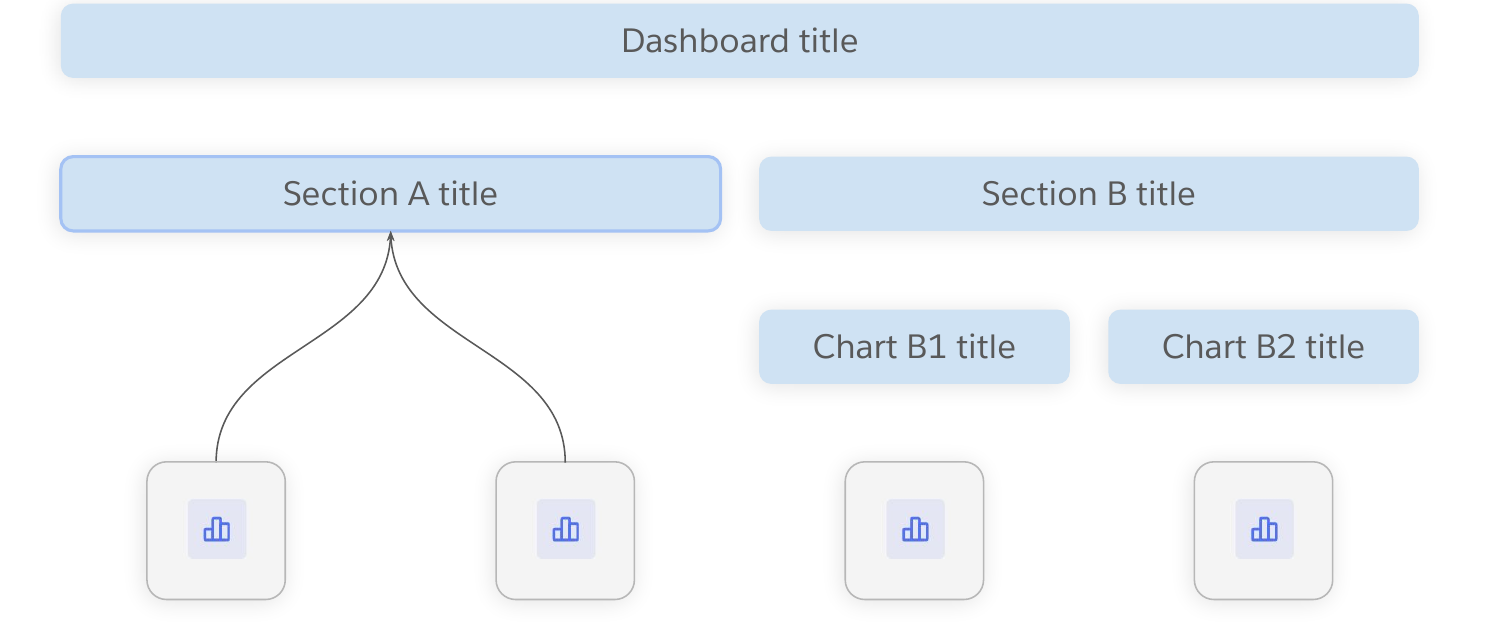}
        \caption{If the children titles are missing, then Section A becomes the leaf node of the dashboard tree and its title gets generated directly from the visualization specifications.}
        \label{fig:tree_b}
    \end{subfigure}

    \begin{subfigure}[b]{\columnwidth}
        \centering
        \includegraphics[width=\columnwidth]{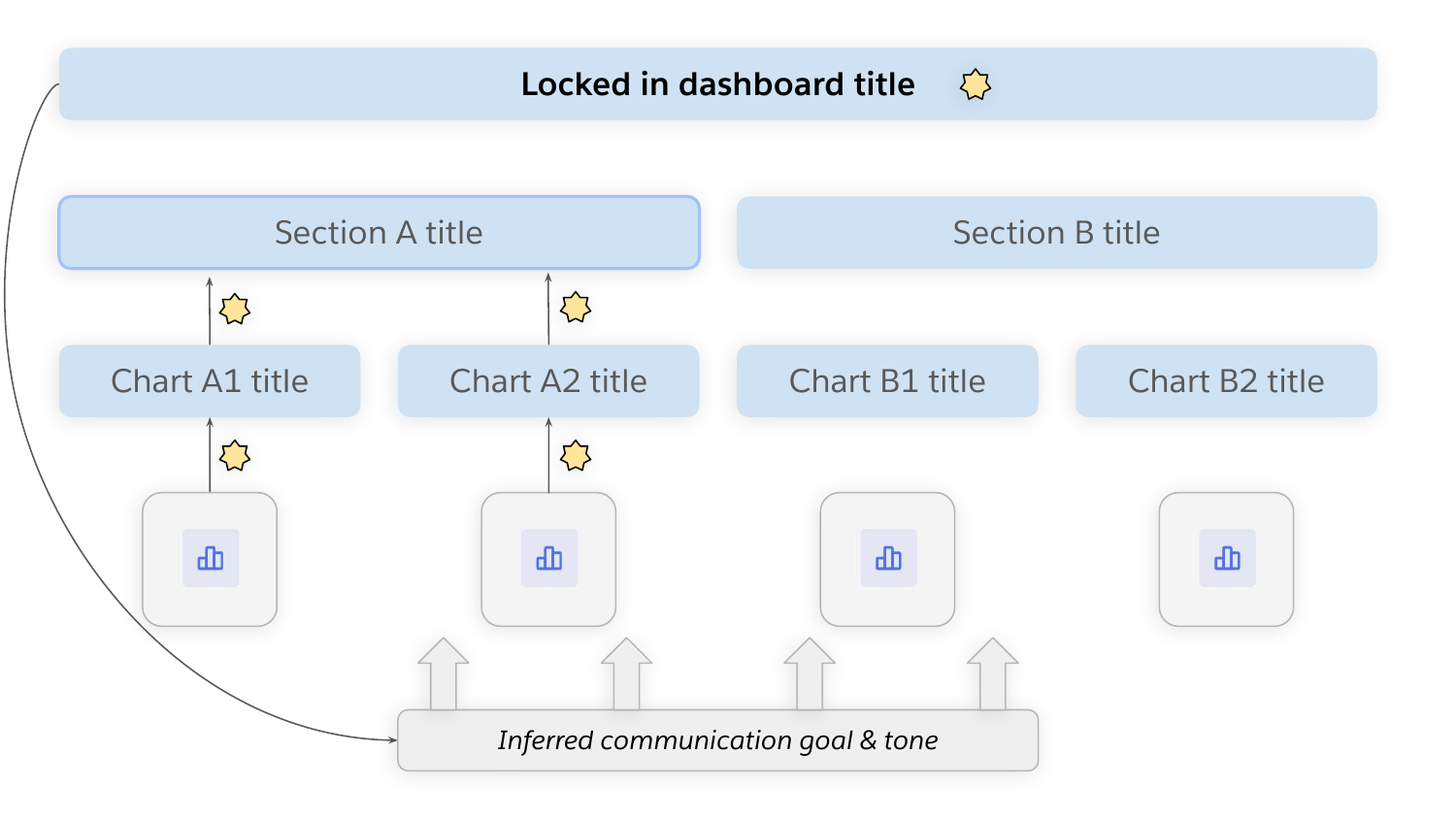}
        \caption{In case the author has manually typed in or locked in any part of the dashboard (indicated by a yellow star), this text will be used by \sys~to infer the author's communication goal and tone. \sys~first collects the locked-in text, and then proceeds to sequentially generate the text bottom-up as before.}
        \label{fig:tree_c}
    \end{subfigure}

    \caption{An illustration of the text generation logic of a section title with respect to the dashboard dependency tree.}
    \Description{Three panels with a schematic tree diagram. In each, the top level is Dashboard title, next level is two section titles, below are four chart titles, and below are the four charts. The information is propagated up from charts.}
    \label{fig:tree}
\end{figure}

\subsection{Text Generation}
\label{sec:generation}

While the user may manually write their own text based on snippet-level suggestions, \sys~also offers an option to press the generation button to have \sys~generate the text. The system generates text by sending a prompt to a large language model (LLM). The prompt consists of (i) an instruction that is specific to each text type, (ii) examples of the text type extracted from the examplar dashboards in our formative work, and (iii) the relevant context from the dashboard. Examples of prompt structure may be found in the supplemental materials.

The instructions are primarily based on an actionable implementation of the text role of the requested text, and the model is asked, for instance, to label the data. The examples provided in the prompt take advantage of the few-shot approach~\cite{brown_language_2020} and give the model a reference for length and tone of language that are commonly found in dashboards. Lastly, the context required by the prompt again depends on the text role and the position of the text in the dashboard frames. We distinguish between four main types of relevant context required by the prompt. 

{\textbf{1. Chart specification}}, which is the code that is used to specify the chart explicitly or implicitly and contains data semantics, data types, transformations, encodings, and interaction specifications. This information is important, for instance, when generating a prompt for text explaining how to interact with a chart, as the system can infer the interactive components or relationships and bindings from the specification.
    
{\textbf{2. Chart visuals}}, or the SVG code of the resulting visualization, allows the LLM to infer what patterns or trends the chart is showing. This information is included in prompts that generate insights or takeaways from the chart. SVG is a useful format for insight identification~\cite{ying2024reviving} as it offers a simple and widely compatible representation of what is actually \emph{seen} in the chart, as opposed to insights inferrable from the data but not observed from the visuals.

{\textbf{3. Downstream text}}, or the text previously written or generated in the children frames, or frames downstream of the current frame in the dashboard tree. If, for a given text role, there exists downstream text, then the prompt will request a summary or comparison of that text rather than generate new text from the charts. This ensures that the generated text (e.g., dashboard takeaway) is consistent with and logically follows the text that comes later in the dashboard (e.g., individual chart insights), avoiding contradictions or non-sequiturs. Figures~\ref{fig:tree_a} and ~\ref{fig:tree_b} show schematic examples of this workflow with and without downstream text, respectively. Another consequence of requiring downstream text for context is that \sys~strictly enforces the asynchronous order of text generation: if multiple snippets across the dashboard are to be generated, the text in the leaf frames is generated first and then passed upwards to its parent frames, with the root frame being generated last.

{\textbf{4. Locked in text}}, which is any text across the dashboard that has been manually edited or marked locked in by the user by pressing a button to the right of the text, will be used in all of the prompts. Since the author had actively written or accepted this text, such text can serve as an indicator of the communication goal of the dashboard, any external context not otherwise inferred from the data that the author would like to add, as well as tone and phrasing. As a result, users can guide \sys's generation by seeding the dashboard with partial text and having the system generate the rest. An example is illustrated in Figure~\ref{fig:tree_c}: while the new text is still generated from the bottom up, locked in text is used to infer the communication goal and remains unchanged.

The dashboard dependency tree plays a crucial role in accurately identifying the relevant context for text generation and touches upon all types of context described above. Figure~\ref{fig:tree} illustrates the flow of dashboard content for text generation across several scenarios. To produce meaningful text, the system maintains information about the content within the text's scope and enforces a strict order of generation: inner text (dependent on charts) is generated before outer text (dependent on the inner text), except for manually edited or locked in text. To ensure consistency throughout the entire dashboard, upstream or sibling text that has been manually entered or locked is used to infer the author's communication goal and is considered in the inner text generation downstream. For example, if a dashboard title highlights a specific insight, the downstream generated text should also support that insight.

\subsection{Implementation}

We implemented \sys~as a web-based browser application and developed the tool using Python and Flask for the back end and React, HTML/CSS, and TypeScript for the front end. We use OpenAI's ChatGPT 4o LLM to generate text. Visualizations in \sys~are specified using Vega-Lite~\cite{satyanarayan2016vega}.

\section{User Study}
\label{sec:study}

We conducted remote user sessions with dashboard creators to evaluate the utility of our proposed workflow for authoring dashboard text and identify useful directions for future work. Participants had the opportunity to use \sys{} in an open-ended authoring task and shared usage impressions and suggestions for future improvement.
Findings suggest that \sys{}'s proposed workflow is flexible, fast, and effective, and inform a number of opportunities for follow-up research. 

\subsection{Methodology}

This section outlines our approach to study design, including how we recruited participants, conducted the evaluation process and the subsequent data analysis.

\subsubsection{Pilots}
We conducted three pilot studies to test our evaluation protocol and identify any issues with the tool: one with a graduate student in visualization and two with professionals experienced in dashboard authoring. Participants were initially tasked with creating a dashboard from scratch until completion. However, this approach proved too time-consuming, as participants continuously refined the dashboards while grappling with unfamiliar data and tools. To address this challenge, we revised the protocol to include a hands-on demo to more quickly familiarize participants with \sys{}'s features. We also developed a more controlled study protocol, allocating a set amount of time---20 minutes---to encourage participants to focus on the workflow and converge towards completion while still exploring the tool’s functionality.

\subsubsection{Recruitment}
For the main study, we recruited 12 professionals ($P1-P12$) with dashboard creation experience from relevant community channels. Participants had, on average, 8.5 years of experience in dashboard authoring (\textit{min}. 4 years, \textit{max}. 18 years) and a diverse range of backgrounds, from data analysts and consultants to data-relevant team management roles. Primary dashboard creation platforms included Tableau and Power BI, but participants also cited a number of support tools for wireframing (e.g., Figma), data management (e.g., Excel, SQL), and programmatic data analysis (e.g., R and Python) as part of their dashboard authoring tool set. 

\subsubsection{Procedure}
The primary goal of our evaluation was to elicit professional dashboard authors' first impressions on \sys{}'s proposed workflow of authoring text. As a result, while providing ample time for the participants to experiment with the tool, we prioritized the discussion and feedback.

Participant sessions were approximately one hour long and entailed three segments. Participants first accessed \sys{} from their own browser and followed a guided walkthrough of its features while sharing their screen. They were then given a dashboard authoring task and up to 20 minutes to complete this task using \sys{} while thinking aloud. After completing the task, participants filled out an adapted System Usability Scale (SUS) questionnaire~\cite{brooke1996sus} and engaged in a discussion about their impressions and experience with \sys{}.

Participants were given one of two authoring tasks: either to (a) create a business-oriented dashboard featuring sales and performance of a fictional store aimed at the ``C-suite" of an investment firm; or to (b) create an infographic-style dashboard featuring interesting highlights from the Olympics, aimed at the general public. The choice of different tasks and data sets allowed us to assess the extent of \sys{}'s support for a broader swath of authoring scenarios and communication goals. Participants were told to consider tasks ``complete'' if they found their dashboard was ``sufficiently described'' and to take no more than 20 minutes. The goal of the time limitation was to ensure that enough time remains to properly walk the participants through the tool's functionality and to solicit feedback and discussion. We found the limitation to provide enough time for participants to give informed feedback on the tool functionality. For the purposes of additionally expediting authoring in the user evaluation, we implemented several preset dashboard layouts, and participants were encouraged (but not enforced) to choose one of the layouts and to consider a diversity of text roles for communication; otherwise, they were free to use \sys{} and craft their dashboards in any way they wished.

Our adapted SUS questionnaire featured all standard SUS questions plus additional items on the perceived extent of control over content generation and text ownership. Interview questions elicited further impressions and suggestions on the various facets of \sys{}, including the text roles, dashboard- and snippet-level guidance, and text generation.

\subsubsection{Data Collection and Analysis}

Sessions were audio- and screen-recorded, and think-aloud impressions and discussions were automatically transcribed and manually vetted. Transcripts were analyzed via a divide-and-conquer thematic analysis approach. First, sessions were equally split between authors, who independently compiled key findings from their subset. The findings were then shared with the group and used to consolidate a rich set of higher-level themes that honored the group's collective interpretations and informed the majority of our qualitative findings.

Text authoring actions with \sys{}---such as creation, editing, generation, refinement, or deletion of text snippets---were also logged and analyzed to uncover use patterns. Together with SUS scores, they provide a quantitative perspective to complement and underscore qualitative findings.

Study instruments, including the session script, results of the modified SUS questionnaire, several examples of resulting dashboards, and the log of participants' text authoring actions during the evaluation, can be found in the supplemental materials.

\begin{figure*}
    \centering
    \includegraphics[width=\textwidth]{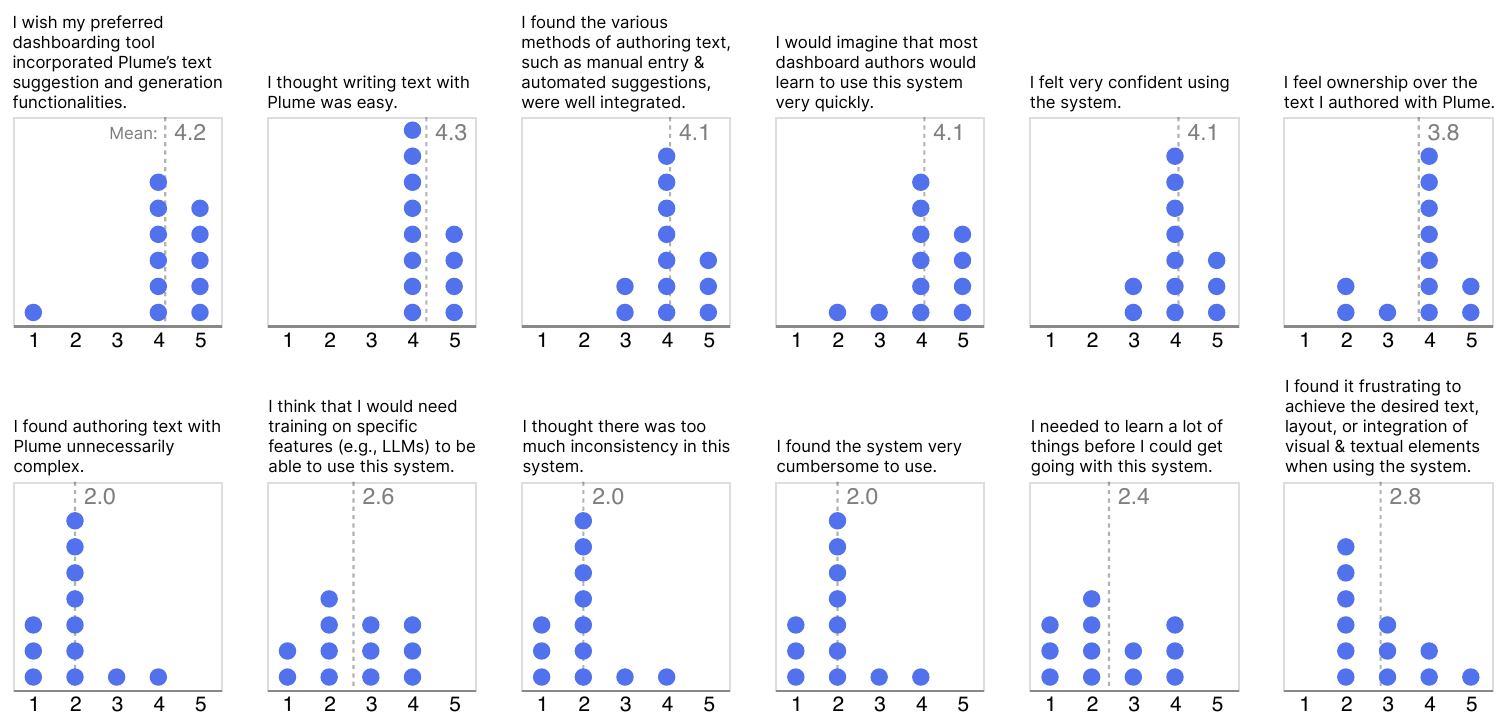}
    \caption{Results of the modified System Usability Scale (SUS) survey. Each circle corresponds to a participant's response. The Vertical dashed line and the number beside the line indicate the mean of Likert scale answers across all participants. The top row represents positive questions (higher is better), and the bottom row shows the results of negative questions (lower is better). The overall SUS score was 72.2, which corresponds to a good level of usability.}
    \Description{A two-by-five grid of dot plots, with 1 through 5 on the x-axis and dots stacked vertically. Overlaid is a vertical line showing the mean of all dots. The top-row answers are all in the range of 3.8 and 4.3. The bottom row means are between 2.0 and 2.8.}
    \label{fig:sus_plot}
\end{figure*}

\subsection{Results and Discussion}
All participants completed the authoring tasks within the expected time range, with a System Usability Scale (SUS) score of 72.2, suggesting good usability and ease of use~\cite{Bangor_SUS_adjective}. Figure~\ref{fig:sus_plot} shows the breakdown of the Likert responses in our SUS survey for each question.

The results from the evaluation of \sys{} revealed several key insights into the tool's effectiveness and workflows, along with areas for improvement based on participant feedback and usage patterns. Throughout this section, we present the insights through a discussion of major themes.

\subsubsection{Text Authoring}
Participants valued \sys{}’s ability to streamline the text authoring process while also supporting new approaches to dashboard creation, with text playing a more central role in the narrative. They appreciated its ability to quickly generate relevant text, overcoming the challenges associated with text authoring, such as `blank canvas paralysis,' where users struggle to start writing from scratch. $P4$ expressed, ``\textit{I don’t really have writer’s block anymore... having a starting point is very helpful.}" Participants also appreciated \sys{}'s ability to generate contextually relevant text based on the data presented. $P3$ emphasized how the tool made ``\textit{contextually relevant word choices}," such as when it generated insights for product sales that aligned with what their manager would want to see. Participants found it helpful to be able to generate text that differentiates across the various text roles within the dashboard. $P8$ stated, ``\textit{The one thing I really like is the differentiation of a lot of the different annotation types. There’s more hierarchy to the way the text is presented.}"

\subsubsection{Integration with Tools and Workflow Strategies}
Participants, in general, found \sys{} to be fast and efficient in helping them generate text relevant for their typical work tasks. $P3$ commented, ``\textit{I think this is absolutely fantastic for ad hoc reporting... You can very quickly do it and create something professional.}" This speed in generating initial text provided users with a strong starting point for refinement, reducing the time spent on the aspects of text authoring they typically find tedious. $P5$ emphasized how much time the tool saved, stating, ``\textit{As a data guy, coming up with words to describe what I'm building can be difficult... auto-generating text and getting a head start is incredibly helpful.}" For participants who regularly authored text for reports, \sys{}'s ability to generate content was a major time saver. $P11$ mentioned, ``\textit{This could be really helpful for the operational dashboards... for the report developer, I would have saved time by just hitting a button to generate instead of me having to go through and individually add in the text boxes.}"

Participants also highlighted the importance of integrating \sys{} with their existing workflows and data exploration tools, many of which are based on existing dashboarding tools like Tableau where charts are created first and then used as building blocks to dashboard composition; we found that the proposed workflow of adding dashboard text fits the way our participants typically author dashboard text. $P7$ stated, ``\textit{I would typically start with laying everything out and then adding informative text}." Similarly, $P8$ commented, ``\textit{Plume is useful in the last mile of this process... maybe I haven't settled on how to tell my story.}" Participants also noted, however, that \sys{}'s text generation features could also be effective if integrated earlier in the analysis process; $P9$ described that this would be especially useful for less experienced analysts: ``\textit{If someone is not very aware of what they like... they will not know how to continue the story. So having your text will help him guide his analysis.}" $P11$ discussed the potential benefits of integrating \sys{} across various reporting workflows, from static reports to communication platforms like Slack and operational dashboards: ``\textit{I see benefits in all three [reporting workflows], especially because it could be dynamic and regenerated.}"

Participants suggested workflow strategies that varied depending on the authoring style. For instance, $P3$ described using a checklist approach, where they systematically went over guided suggestions, added placeholders, and generated text.  $P8$ noted that the tool prompted them to think more critically about their captions and footnotes: "\textit{It’s already having me think more critically about how I have a lot of my captions and footnotes set up.}" $P1$ and $P4$ experimented with a maximalist approach, adding all placeholders before removing unwanted snippets, whereas $P2$ employed a build-up approach, intentionally adding snippets one by one. 

Analyzing the log of actions during the evaluation revealed that participants used \sys~to create, on average, 38 text snippets (SD 14.4, min 20, max 69), and removed 16.4\% (SD 17.0\%, min 0.0\%, max 60\%) of them.

\subsubsection{Customization and Control}
\sys{} was valued for generating an initial set of text that authors could refine and customize. $P6$ noted, ``\textit{This is creating a starting point because maybe you didn't realize what you could or should do.}" The tool allowed authors to quickly iterate on the generated content, making it easier to start from something instead of from scratch. $P2$ echoed this sentiment, saying, ``\textit{Would I use this text? Maybe not, but I think this is a good starting point that would prompt me to think about what to put here.}" 

Participants also noted \sys{}'s potential to encourage exploration and assess different narrative options. $P3$ found the tool effective for summarizing context rapidly, stating, ``\textit{Plume would be great for wireframing; it lets me quickly see how a narrative flows.}" This flexibility allowed authors to iterate on different text roles and narrative structures. However, $P1$ and $P4$ wanted more control over text metrics, particularly indicators like word count, readability, and tone, with the desire to be able to adjust these more granularly: ``\textit{If I could see sliders for word count and tone, I’d feel more in control of the text it’s generating}. [$P1$]"

Participants' action log analysis revealed that they accepted, on average, 66.1\% of snippets created by \sys~as-is (SD 21.4\%, min 20.0\%, max 88.2\%). Of the rest, participants manually edited 8.7\% (SD 4.7\%, min 0\%, max 16.7\%) of snippets, used \sys~to shorten or simplify 8.5\% (SD 5.8\%, min 2.2\%, max 20.8\%), and asked \sys~to completely regenerate 12.4\% (SD 7.5\%, min 2.3\%, 25.0\%).

\subsubsection{Dynamic Updates and Automation}
Participants expressed enthusiasm about \sys{}'s ability to automatically generate new text based on data changes, aligning the text with the dynamic nature of dashboards. $P7$ mentioned, ``\textit{All of our dashboards at work are connected to live data, and that's kind of always the problem. I never want to put static text in a dashboard because trends will change, and then I'd have to go in and change the dashboard.}" 

However, one of the challenges discussed was balancing automation with manual control, particularly around real-time updates. $P7$ again reflected on this topic and expressed hesitancy, saying, ``\textit{I'd still want to vet it to make sure it wasn't saying something that was not accurate.}'' Indeed, during our evaluations, participants edited or deleted about a third of text snippets suggested by \sys{}. This highlights the tension between automation and user control: users appreciate the speed and efficiency of real-time updates but also want to ensure that the content remains accurate, especially in high-stakes contexts. $P12$ further emphasized the challenge of keeping AI-generated text aligned with dynamic data, stating, ``\textit{If this was Superstore, and we have volatile sales patterns, is this auto-generating or is this fixed?}'' The potential for stale or misleading text due to automatic updates was a concern across multiple participants, prompting a need for mechanisms that provide notifications for updates or a review process before changes go live.

Participants also suggested that dynamic updates should be more customizable. $P3$ proposed the ability to mark specific text phrases for regeneration, allowing for finer control over which parts of the dashboard text are updated. This approach could enable a balance between the convenience of automation and the assurance of accuracy through human review---\textit{``People are happier to correct than create.''}

\subsubsection{AI-Generated Text, Trust, and Usability}
Participants often found the text generated by \sys{} more verbose than they intended. $P8$ commented, ``\textit{This will be a lot. I might get rid of some of these}," highlighting the need to pare down the output. On average, we observed that users employed the system’s LLM-supported refinement and manual text editing at nearly the same rate, with each method accounting for 8.5\% and 8.7\% of all snippets, respectively. While some participants appreciated the speed and simplicity of automatic refinement, others, like $P1$, felt that the tool's `shorten' and `simplify' functions did not offer the control they needed. Since the functionality fell short of their expectations, $P1$---the only participant who indicated in the survey that they would not like to incorporate {\sys} into their preferred tools---opted for manual edits to better manage the text’s length and tone.

While \sys{}'s ability to generate text based on connected data was seen as a major advantage, as discussed before, concerns about inaccuracies within these dynamic updates reflect participants' hesitation to blindly let AI generate all of their dashboard text. $P6$ noted, ``\textit{I love that it can update based on the data, but I don't fully trust it to be correct every time without reviewing}." Participants suggested that notifications for updates could help ensure that authors maintain control over the accuracy of the text. Additionally, $P4$ noted, ``\textit{Notifications, when text gets stale or outdated, would be useful, especially if data changes monthly or hits certain thresholds}." To address these concerns, participants suggested more guardrails and flexibility in the tool. $P4$ recommended generating multiple alternatives for each snippet, allowing users to ``pick the best" from a set of generated options. $P1$ echoed this by calling for more control over the tone, particularly for insights text role, where a more precise tone is often required to carefully craft a takeaway.

\section{Future Implications and Limitations}

In this section, we discuss limitations and opportunities for future improvements of assisted text authorship in dashboards specifically, as well as broader implications for the research community.

While our user study provided valuable insights into dashboard authors' initial impressions of \sys, as well as their attitudes and trust toward generative AI in dashboard text, it is crucial to explore its impact on longer-term workflows~\cite{jacovi:2021}. As tools like \sys~are adopted in practice, future research should investigate how data workers and dashboard authors integrate these systems into their everyday processes, particularly when working with their own data sets and dashboards. Longitudinal studies would be essential to observe more complex insights about how \sys~is integrated into existing workflows: how trust in generated text evolves over time, how \sys~influences dashboard creation in real-world settings, and which types of dashboards benefit most---or least---from this assistance. \rrr{Further studies on real-world adoption could also highlight how users at different levels of expertise engage with {\sys} in distinct ways. While all participants in our study were experienced dashboard authors, they speculated that novice users might find the most value in assistance with generating initial ideas and establishing their unique voice.}

In addition to studying how \sys~integrates into long-term workflows, future work should also investigate how to better collect and leverage analysts' implicit knowledge about the data throughout the analysis process to support more effective dashboard text generation. Analysts often accumulate critical insights, metadata, and observations during their exploration of the data---long before they begin writing text for dashboards. Capturing this context earlier in the workflow and even surfacing it to the user~\cite{lin_data_2023} could enable systems like \sys~to generate more accurate, relevant, and nuanced text that aligns closely with the analyst’s intentions and the data set’s nuances. Future research could focus on how to seamlessly integrate mechanisms for annotating and preserving contextual information at key stages of the analysis, such as during data wrangling, chart creation, or hypothesis formation. By doing so, \sys~could anticipate the needs of dashboard authors, offering richer suggestions for labels, insights, and contextual explanations that are grounded in the specific analytical journey of the user.

\sys's hierarchical structure and editing features help users maintain consistency in their dashboard text. However, there are still opportunities to enhance the coherence of the generated content. For instance, we observed that the same hex colors might be referred to by different terms, such as ``navy blue" in one section and ``dark blue" in another. Similarly, similar sections of the dashboard could end up with slightly different titles due to the inherent variability of LLM output. Future work could address these issues by fine-tuning LLM prompts, implementing stricter text generation templates, and offering more explicit user controls for reviewing and synchronizing text across related sections. Additionally, the clear delineation of text roles in \sys{} enables future work to enhance the performance of each role independently. For instance, \rrr{although we tested our system to confirm that it performs well on common Vega-Lite charts (our evaluations used line charts with and without uncertainty bands, bar charts, scatter plots, and area charts),} computational frameworks specifically fine-tuned for extracting insights based on the data {~\cite{tang2017extracting, ying2024reviving}} \rrr{could be added in a modular way, improving the \emph{insight} text generation in general and for more complex chart types in particular}.

Another area of exploration lies in expanding \sys{}’s integration into collaborative workflows. Tools like \sys{} could be integrated into platforms such as Slack, Teams, and email, allowing users to easily retarget dashboard content~\cite{kim_bringing_2024} or generate and share automated summaries or key insights in real time. For instance, \sys{} could automatically send concise updates whenever a key metric changes, delivering essential information without requiring users to leave their communication platform. This scheduled reporting functionality, embedded within messaging platforms, could further extend its utility for operational dashboards and reporting workflows.

\sloppy
Moreover, there is growing potential for exploring text-to-voice functionalities to enhance the multimodal interaction of dashboards~\cite{stokes:2024}. By modulating prosody and acoustic characteristics based on the type of text (e.g., annotations vs. insights), users could receive auditory feedback that aligns with the text’s role in the dashboard, allowing for a more dynamic and engaging user experience. This capability could be particularly useful in time-sensitive environments where real-time updates could be delivered as automated voice alerts.

Multilingual text generation also offers significant opportunities for expanding \sys{}'s capabilities, allowing global teams to interact with dashboards in their preferred languages. Visualization descriptions and interpretations often vary across different languages, as well as due to cultural differences in how a language is perceived and used~\cite{rakotondravony2022probablement, rakotondravony2023beyond}. Future work could explore how the system could provide role-specific translations that maintain the intended clarity and structure, supporting broader collaboration across linguistic barriers.

Finally, a key challenge for future research will be finding the right balance between real-time updates of dashboard text and analysts' control over content. Tools like \sys~open the possibility of connecting text directly to live data feeds, allowing the text to update dynamically as data changes. While this functionality could help resolve the issue of outdated textual descriptions, fully automating text updates raises concerns about accuracy and trust, especially since text---unlike charts---can be more interpretive or imprecise. This tension underscores broader questions about the supervision of AI-generated content, as analysts may need to review and vet textual updates more carefully to ensure they remain aligned with the data. Future research will need to explore how to manage this trade-off, possibly by developing systems that allow selective automation, where some text updates are handled in real time while others require analyst approval. Such advancements will have significant implications for how AI tools integrate into workflows, balancing efficiency with the need for careful oversight.
\section{Conclusion}
We present \sys{}, a system designed to explore and operationalize the design space of text roles in dashboards, addressing the need for clear, contextually appropriate text that enhances data-driven communication. The tool scaffolds the composition of text across different semantic levels, providing support for a range of text roles such as labels, insights, and contextual descriptions. By integrating AI-generated suggestions with human-in-the-loop oversight, \sys{} bridges the gap between the rich visualizations typical of dashboards and the equally important textual elements that guide interpretation and understanding. A preliminary evaluation of the tool highlights dashboard authors' excitement about text composition assistance and the potential for future research to refine text generation workflows, especially in balancing the trade-offs between AI-driven automation and manual control for effective communication. These findings reveal opportunities for improving tools that cater to the diverse needs of dashboard authors, enabling more coherent and contextually appropriate data-driven insights. By drawing on writer Robert Louis Stevenson's quote, \textit{``The difficulty of literature is not to write, but to write what you mean; not to affect your reader, but to affect him precisely as you wish."} \sys{} seeks to bridge the gap between the author's intention and the audience's understanding, facilitating more purposeful data-driven communication.


\begin{acks}
We thank the members of the Tableau community and the Tableau Research team for their valuable feedback and insights, which have helped shape and refine this work.
\end{acks}

\bibliographystyle{ACM-Reference-Format}
\bibliography{paper-2024-plume}

\end{document}